\documentclass{pr-imfp00}

\newcommand{\bibit}{\nineit}
\newcommand{\bibbf}{\ninebf}

\font\nineit=cmti9
\font\ninebf=cmbx9

\newcommand{\bi}{\begin{itemize}}
\newcommand{\ei}{\end{itemize}}
\newcommand{\text}[1]{\mbox{#1}}
\newcommand{\E}[1]{10$^{#1}$~eV}       
\begin{document}

\title{Theoretical and Experimental Topics on Ultra High Energy Cosmic Rays}

\author{A. LETESSIER-SELVON}

\address{LPNHE, Universities of Paris 6 \& 7,  IN2P3-CNRS,\\
 Tour 33 RdC 4 place Jussieu,\\
 75252 Paris Cedex 05, France\\
E-mail: Antoine.Letessier-Selvon@in2p3.fr}

\maketitle

\abstracts{Since their first observation in 1962, the existence of
Ultra High Energy Cosmic Rays (UHECR) remains a mystery in modern astrophysics.
Those cosmic rays, with energies well above 50 EeV ($50\times 10^{18}$eV), can hardly 
be accelerated, even in the most active parts of our universe such as FR-II radio galaxies or AGNs, 
nor can they travel on distances larger than 100 Mpc.  In the following some of the production and 
acceleration models for UHECR are reviewed and some of the transport issues are exposed. 
Finally the detection and identification on Earth of those ``cosmic bullets'' are presented.}

\section{Introduction} \label{intro}
\noindent 
This lecture is mainly concerned with the problems of the existence
and observation of cosmic rays whose energies are above
$5\times10^{19}$ eV. Such cosmic rays -~for which we shall use the term
``ultra  high energy cosmic rays" or UHECR~- are exceptional for the
following reasons:
\bi
\item They are above the Greisen-Zatsepin-Kuzmin (GZK) cutoff\cite{Greisen} which corresponds to
the proton energy threshold for pion photo-production on the cosmic microwave background
(CMB). Similar cutoffs exist at lower energies for gammas interacting with
background photons (CMB, infra-red or radio waves). Consequently, and except for neutrinos, if the UHECR
observed on Earth are due to the known stable particles, they must be produced in our vicinity.
At the GZK cutoff, the ``visible" universe shrinks suddenly to a sphere of a few tens of 
megaparsecs (Mpc).
\item There are very few conventional astrophysical sources able to accelerate particles at
energies exceeding \E{20}, those of the most energetic UHECR that have been
observed up to now. 
\item At such energies and in most field models, the bending effect of the galactic and extragalactic 
magnetic fields are quite weak. Thus, even for charged particles,
the reconstructed incident direction points toward the source within
a few degrees. Unlike with lower energy cosmic rays one can do point-source-search astronomy
with UHECR.
\ei
The widely shared excitement about the UHECR comes from the
above considerations and from the study of the scarce data available. 
If the sources are astrophysical macroscopic objects,
they must be visible through some counterpart that optical or radio-astronomy
should detect. But there are no remarkable object visible in the directions pointed at by
UHECR. There is even no convincing evidence that one can find any
correlation between the incoming directions and the inhomogeneous distribution
of matter in our vicinity. 

In the following, we shall develop in detail the facts and arguments briefly
mentioned in this introduction.
 In section 2 we present the candidate source characteristics and we discuss the transport problems.
Section 3 is devoted to cosmic ray interactions in the atmosphere and section 4 to detection 
techniques. Finally section 5 describes some of the available experimental results.

To avoid repetitive use of large powers of ten, the energy units in the
following will mostly be in zetta-electron-volts (ZeV, \E{21}) and
exa-electron-volts (EeV, i.e. \E{18}).

\section{Production and Transport of UHECR}\noindent
Today's understanding of the phenomena responsible for the production of UHECR, 
i.e. the transfer of macroscopic amounts of energy to 
microscopic particles, is still limited.
One distinguishes two classes of processes: the so called ``Top-Down''
and ``Bottom-Up'' scenarios. In the former,
the cosmic ray is one of the stable decay products of a 
super-massive particle. Such particles 
with masses exceeding 1~ZeV can either be meta-stable relics of some primordial field or highly 
unstable particles produced by the radiation, interaction or collapse of topological defects. 
Those processes are reviewed in Section~\ref{TopDown}

In the second scenario discussed in section~\ref{BottomUp} the energy is transferred 
to the cosmic rays through their interaction with electromagnetic
fields. This classical approach does not require new physics as opposed to the ``Top-Down'' mechanism, but does not exclude it either since, in some models, the accelerated particle - the cosmic ray -
is itself ``exotic''. 

Once accelerated the cosmic rays must propagate from their source to the observer. 
At energies above 10~EeV and except for neutrinos, the Universe 
is not transparent to \underline{ordinary} stable particles on scales much larger than about 10 Mpc. 
Regardless of their nature, cosmic rays lose energy in their interaction with the various photon 
backgrounds, 
dominantly the copious Cosmic Microwave Background (CMB) but also the Infra-Red/Optical (IR/O) and 
the Radio backgrounds. 
The GZK cutoff puts severe 
constraints on the distance that a cosmic ray can travel before losing most of its energy or being 
absorbed. The absence of prominent visible astrophysical objects in the direction of the 
observed highest 
energy cosmic rays together with this distance cutoff adds even more constraints on the ``classical'' 
Bottom-Up picture.

It is beyond the scope of this lecture to describe all the scenarios - they are
far too numerous - proposed for the production of the UHECR. 
Let us simply agree on the fact that the profusion of models shows that none of them is totally 
satisfactory and that data are not very constraining. 
Consequently we will try to present, from an experimentalist's point of view, 
the main features of the various categories of models.
We will also try to focus on the possible experimental constraints, if they exist, or on the 
problems related to the UHECR and which remain unsolved. For a more detailed review we urge the 
reader to
consult the excellent report by P.Bhattacharjee and G.Sigl\cite{Batt_Sigl} and the references therein. 
Extensive use of this report is made in some of the following sections where we avoided repeated 
reference to it. 

At first sight, it would seem natural to discuss potential sources and acceleration mechanisms before
the description of the cosmic ray transportation to Earth. However, and because the attenuation or 
interaction lengths are relatively
short and strongly energy dependent in the range of interest, the observed spectra do not only depend on the
nature of the sources  but also on their distribution. In addition, the GZK cutoff 
puts important constraints which we prefer to discuss before describing the possible 
nature of the sources themselves.

\subsection{Propagation}\noindent
We will focus here on the propagation of atomic nuclei (in particular protons) and photons. 
Electrons are not considered as potential UHECR because they radiate most of their energy while 
crossing the cosmic magnetic fields.  Among the known stable particles, and within the framework of the Standard Model, those are
the only possible candidates for UHECR. As mentioned in Section~\ref{Compo}, the 
actual data effectively favor a hadronic composition. 

Neutrinos and the lightest super-symmetric particles (LSP) should deserve special attention as they may travel
through space unaffected even on large distances.
However, for neutrinos the interaction should occur uniformly in atmospheric depth,   
a feature which is not reproduced by the current data. 
While neutrinos may very well be one of the components of the high energy end of the cosmic ray 
spectrum and prove to be an unambiguous signature of the new physics underlying the 
production mechanisms (see below) they do not seem to dominate the observations at least up 
to energies of a few \E{20}. 

The LSP are expected to have smaller interaction cross sections with photons and a higher threshold for pion photo-production due to their higher 
mass (see Eq.~(\ref{eq:GZK}) below). Therefore they may travel unaffected by the CMB on distances 
10 to 30 times larger than nucleons. However in usual models the LSP is neutral and cannot be 
accelerated in a Bottom-Up scenario and must be produced as a secondary of an accelerated charged 
particle (e.g. protons). This accelerated particle must reach energies 
at least one order of magnitude larger than the detected energy (order of ZeV) and will produce photons. 
The acceleration site should therefore be detectable as a 
very powerful gamma ray source in the GeV range. In a Top-Down scenario including Super-Symmetry, 
the problem of propagation is of somewhat lesser importance as the decaying super-massive particles 
may be distributed on cosmological or on nearby scales and are, in any case, 
invisible (see Section~\ref{TopDown}). Finally let us stress that the analysis of the 
UHECR shower shape limits the mass 
of the cosmic ray to about 50~GeV,\cite{Albuquerque} an additional  constraint for the LSP candidate.
\subsubsection{Protons and Nuclei: The GZK cutoff}\label{Toto}\noindent 
The  Greisen-Zatsepin-Kuzmin (GZK) cutoff (see Section~\ref{intro}) threshold for  
collisions between the cosmic microwave background (CMB) and protons (pion photo-production) 
can be expressed in the CMB ``rest'' frame as 

\begin{equation}\label{eq:GZK}
 E_{th} \simeq \frac{E_\gamma^{lab} m_p}{2\epsilon} \sim \frac{7\times 10^{16}}{\epsilon}eV
\end{equation}
where $\epsilon$ is the CMB photon energy, 
$E_\gamma^{lab} = m_\pi^2/2m_p + m_\pi$ is the photon threshold for a proton at rest 
and $E_{th}$ the proton threshold in the CMB frame, all in electron-volts. 
For an energetic CMB photon with $\epsilon=$ \E{-3}, $E_{th}$ is $7 \times $\E{19} which is where one expects the GZK 
cutoff to start.
 
The interaction length for this process can be estimated from the pion photo-production cross section (taken beyond the $\Delta$
resonance production) and the CMB photon density:
$$ L = (\sigma \rho)^{-1} \simeq 1.8\times10^{25}\,\text{cm} \simeq 6\, \text{Mpc} $$        
for $\rho = 410\,\text{cm}^{-3}$ and $\sigma = 135~\mu$barns. 

\begin{figure}[!htb]
\vspace*{13pt}
\begin{center}  
\epsfig{file=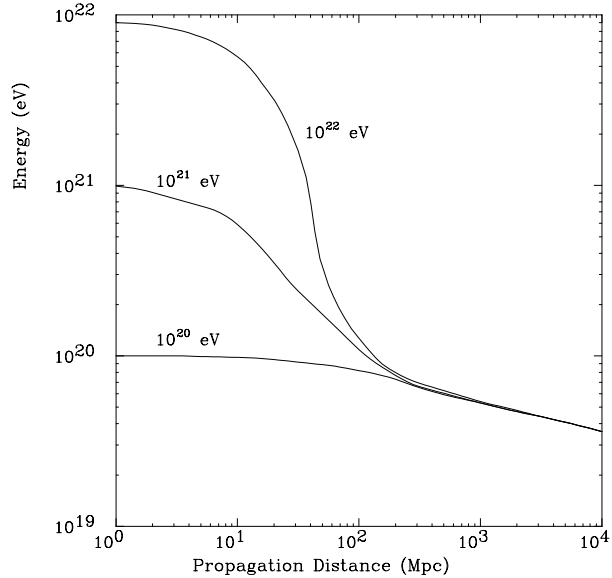, width = 8cm}
\end{center}
\vspace*{13pt}
\caption{Energy of a proton as a function of the propagation distance through the 2.7K cosmic 
background radiation for various initial energies.\label{adrf25}} 
\end{figure}

The energy loss of protons of various initial energies as a function of the propagation 
distance is shown in Figure~\ref{adrf25}. Above 100 Mpc the observed energy is below 
\E{20} regardless of its initial value. One should point 
out that this reduction is not the consequence of a single catastrophic process but of many 
collisions (more than 10) 
each of which reduces the incident energy by 10 to 20\%.
Therefore the probability to travel without losses is negligible.

\par
A proton may also produce $e^+e^-$ pairs on the CMB at a much lower threshold (around $5\times$\E{17}) 
but the cross section is orders of magnitude smaller and together with a much smaller energy loss per interaction
the overall attenuation length stays around 1~Gpc.

\begin{figure}[!htb]
\vspace*{13pt}
\begin{center}  
\epsfig{file=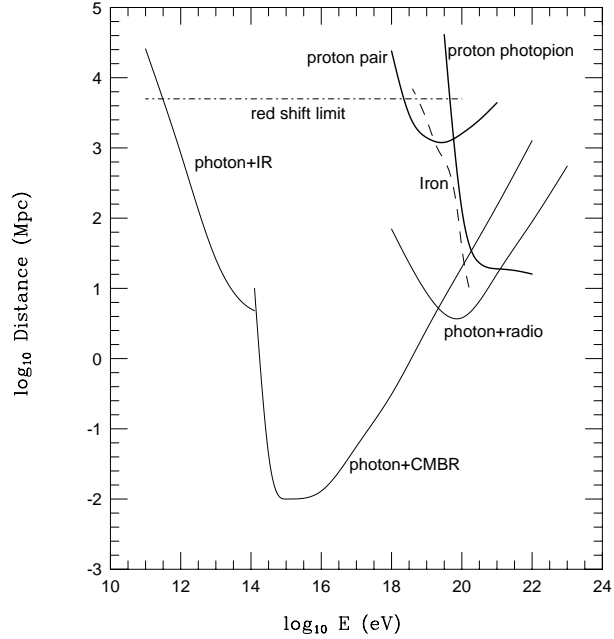, width = 8cm}
\end{center}
\vspace*{13pt}
\caption{Attenuation length of photons, protons and iron in various background radiation as a function of 
energy. The dot-dashed line represents the absolute upper limit on the distance a 
particle can travel toward Earth, regardless of its initial energy.\label{adrf24}} 
\end{figure}

For nuclei, the situation is in general more difficult. 
They undergo photo-disintegration in the CMB and infrared radiations
losing on average 3 to 4 nucleons per Mpc when their energy exceeds 2$\times$\E{19} to 2$\times$\E{20} depending on the IR background density value. 
The IR background is much less well known than the CMB and the attenuation length 
(see Figure~\ref{adrf24}) derived for nuclei must be taken with precaution. 

\subsubsection{Electrons and Photons: Electromagnetic Cascades\label{photon}}\noindent
Top-Down production mechanisms predict that, at the source, photons (and neutrinos) 
dominate over ordinary hadrons
by about a factor of ten. An observed dominance of gammas in the supra-GZK range would then be an almost 
inescapable signature of a super-heavy particle decay. Photons are also secondaries of more 
ordinary processes such as pion photo-production; their propagation is thus worth studying.
Unlike photons, electrons and positrons cannot constitute the  primary CR as the radiation energy losses they 
undergo forbid them to reach the highest energies by many orders of magnitude. 

High energy photons traveling through the Universe produce $e^+e^-$ pairs when colliding with the 
Infra-Red/Optical (IR/O), CMB or Universal Radio Background (URB) photons. As can be seen on 
Figure~\ref{adrf24} the attenuation length gets below 100 Mpc for photon energies between $3\times$\E{12} 
and \E{22}. In this energy range, nearly 10 orders of magnitude, the Universe is opaque
to photons on cosmological scales.

Once the photon converted, the $e^+e^-$ pair will in turn produce photons mostly via Inverse Compton Scattering (ICS) 
(the case of synchrotron radiation, usually non dominant, will be treated in the next section). At our energies,
those two dominant processes  are responsible for the production of electromagnetic (EM) cascades.
  
Far above the pair production threshold ($s \gg 4 m_e^2$, where $\sqrt{s}$ is the CM energy) 
the ICS ($\sigma_{\text{ICS}}$) and the pair production ($\sigma_{\text{pp}}$) 
cross sections are related by~:
$$ \sigma_{\text{pp}} \approx 2 \sigma_{\text{ICS}} = \frac{3}{2} \sigma_T \frac{m_e^2}{s} \log\left ( \frac{s}{2m_e^2}\right )$$
where $m_e$ is the electron mass and 
$\sigma_T=8\pi\alpha^2/3m_e^2=665$~mbarn the Thomson cross section for photon
elastic scattering on an electron at rest. The $1/s$ dependence implies that far from the pair  
threshold the EM cascade develops slowly as it is the case when the initial photon energy is above \E{22}. 

At the pair production threshold ($s \sim 4 m_e^2$), the pair cross section reaches \mbox{$\sim170$~mbarn} and 
$\sigma_{\text{ICS}}$  is nearly equal to the Thomson cross section. The EM cascades develop very rapidly. From 
Figure ~\ref{adrf24} one sees that at the pair production threshold on the CMB photons 
($2\times$\E{14}) conversion occurs on distances of about 10~kpc (a thousand times smaller than for protons at 
GZK energies) while subsequent ICS of electrons on the 
CMB in the Thomson regime will occur on even smaller scales (1~kpc). 

As a consequence, all photons of high energy (but below \E{22}) will produce, through successive collisions
on the various photon backgrounds (URB, CMB, IR/O), lower and lower energy cascades 
and pile up in the form of a diffuse 
photon background below \E{12} with a typical  power law spectrum of index $\alpha=1.5$.  This is a very important fact
as measurements of the diffuse gamma ray background in the $10^7$-\E{11} range done for example by EGRET\cite{EGRET} 
will impose limits on the photon production fluxes of Top-Down mechanisms and consequently on the abundance of 
topological defects or relic super-heavy particles.

\subsubsection{Charged Particles: Magnetic Fields}\noindent
The effect of magnetic fields (galactic or extragalactic) on the deflection of charged particles will be 
reviewed in Section~\ref{MagField}. Here we will present some of the effects of the fields on EM cascade 
production.

Electrons and positrons produced through EM cascades lose energy via synchrotron radiation at a rate given by:

$$ -\frac{dE}{dt} = \frac{4\alpha^2}{3 m_e^2}<B^2>\left(\frac{E}{m_e}\right)^2,$$ 
where we assume a random field $B$ isotropically distributed with respect to the 
electron direction.

At high enough energy, i.e.
$$ E \sim \left(\frac{B}{10^{-9}\,\text{G}} \right)^{-1}10^{19}\,\text{eV}$$
this process will dominate over ICS on URB or CMB photons. In a nano\-Gauss field and at \E{19} 
the loss is about $3\times$\E{18} over 100 kpc.
The above threshold is not very strict as it depends on the 
URB density which is not a very well known quantity. The emitted gammas have a typical energy given by\cite{Batt_Sigl}
$$E_{\text{synch}} = 6.3\times 10^{11}\left(\frac{E}{10^{19}\,\text{eV}}\right )
\left ( \frac{B}{10^{-9}\,\text{G}}\right )\text{eV}.$$
Again low energy photon flux measurements will put constraints on the extragalactic fields and/or on the initial photon 
flux. 

Above threshold, the synchrotron radiation will damp the electron-positron pair energy 
extremely quickly. At 100~EeV in a $10^{-9}$~G magnetic field the attenuation length is of the order of 20~kpc. 
If one observes gammas above \E{20} they could not be high energy secondaries (e.g. from ICS) 
of an even higher energy photon converted into a pair. They must instead be primary ones. 
Consequently, their flux $j_\gamma(E)$ per unit area and unit solid 
angle at a given energy is directly related to the source distribution without any transport nor cosmological effects in between~:
$$ j_\gamma(E)
\sim \frac{1}{4\pi}l_\gamma(E)\phi(E)$$
where $\phi(E)$ is the source density per unit time and energy interval and $l_\gamma(E)$ is the 
photon interaction length.  

Of course quantitative predictions of such effects is pending definite measurements of the galactic and
extragalactic magnetic fields. Although the magnetic fields of the galactic disc 
are now believed to be fairly well known 
this is not the case of the ones in the halo or extragalactic media. As mentioned in 
Section~\ref{Xfield}, several authors
advocate our bad knowledge of those fields in explaining the puzzling observational data and question both 
the typical value of $10^{-9}$~G and the coherence length 
of 1~Mpc usually assumed for the extragalactic fields.\cite{Sigl_Lemoine_Biermann}

\subsection{Conventional acceleration: Bottom-Up scenarios}\label{BottomUp}\noindent
One essentially distinguishes two types of acceleration mechanisms~:
\begin{itemize}
\item Direct, one-shot acceleration by very high electric fields. This occurs in or near 
very compact objects such as highly magnetized neutron stars or the accretion disks of black holes. 
However, this type of mechanism does not naturally provide a power-law spectrum. 

\item Diffusive, stochastic shock acceleration in magnetized plasma clouds which generally 
occurs in all systems where shock waves are present such as supernova remnants or radio galaxy hot spots. 
This statistical acceleration is known as the Fermi mechanism of first  (or second) order, depending on whether
the energy gain is proportional to the first (or second) power of $\beta$, the shock velocity.   
\end{itemize} 
Extensive reviews of acceleration mechanisms exist in the literature, e.g. on acceleration 
by neutron stars,\cite{Olinto2}  shock acceleration and propagation,\cite{Protheroe}  
non relativistic shocks,\cite{Drury}  and relativistic shocks.\cite{Kirk_Duffy}

\begin{figure}[!htb]
\vspace*{13pt}
\begin{center}  
\epsfig{file=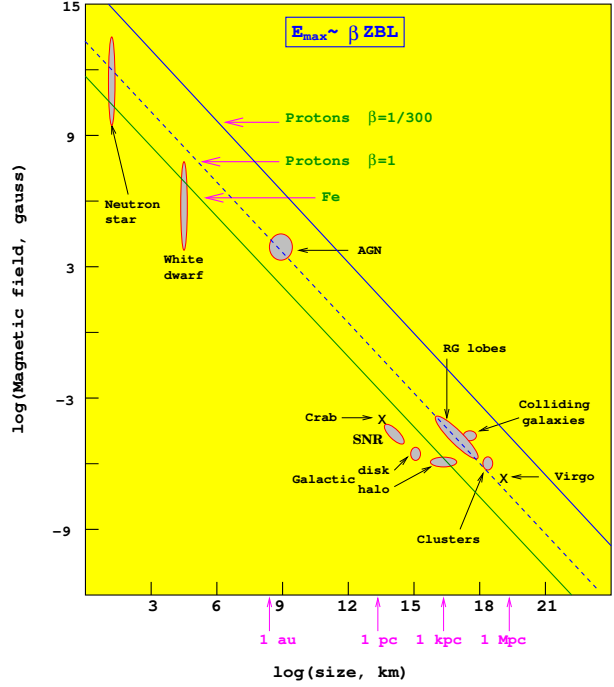, width = 8cm}
\end{center}
\vspace*{13pt}
\caption{Size and magnetic field strength of possible acceleration sites. Objects below the diagonal 
lines cannot accelerate the corresponding elements (Iron with $\beta=1$ or protons $\beta=1$ and 
$\beta=1/300$) above \E{20}.\label{Hillas-Diagram}}
\end{figure}

Hillas has shown\cite{Hillas} that irrespective of the details of the acceleration 
mechanisms, the maximum energy of a particle of charge $Ze$ within a given site of size $R$ is: 
\begin{equation}
E_{\text{max}}\approx\beta Z\left(\frac{B}{1\,\mu \text{G}}\right )\left(\frac{R}{1\,\text{kpc}}\right)10^{18}~\text{eV}
\label{eq:Hillas}
\end{equation}
where $B$ is the magnetic field inside the acceleration volume and $\beta$ the velocity of the shock wave 
or the efficiency of the acceleration mechanism. This condition essentially states that the Larmor radius of the 
accelerated particle must be smaller than the size of the acceleration region,
and is nicely represented in the Hillas diagram shown in 
Figure \ref{Hillas-Diagram}.   

\subsubsection{Candidate sites}\noindent
Inspecting the Hillas diagram one sees that only a few astrophysical sources satisfy the necessary, but not sufficient, 
condition given by Eq.~(\ref{eq:Hillas}). Some of them are reviewed e.g. by Biermann.\cite{Biermann} Let us 
just mention, among the possible candidates, pulsars, Active Galactic Nuclei 
(AGN), Fanaroff-Riley Class~II (FR-II) radio galaxies and Gamma Ray Bursts (GRB). 

\begin{quote}
\par {\it Pulsars}~\\
From a dimensional analysis, the electric field potential drop in a rotating magnetic pulsar is given by:
\begin{equation}
\Delta \Phi = \frac{B\times R^2}{\Delta T} 
\label{eq:pulsar}
\end{equation} 
One obtains $e\Delta \Phi=100$~EeV with $B=10^{9}\,$T, $\Delta T=10^{-3}\,$s and $R=10^4\,$m. 
However the high radiation density in the vicinity of the pulsar will produce  
$e^+e^-$ pairs from conversion in the intense magnetic field.\cite{Olinto2} 
These pairs will drift in opposite directions along the field lines and 
short circuit the potential drop down to values of about \E{13}. Moreover in the above dimensional analysis 
a perfect geometry is assumed. Actually, a more realistic geometry would introduce an 
additional factor $R/c\Delta T\sim 0.1$,  and further decrease the initial estimate. 
Finally, as will be described in the next section, synchrotron radiation losses in such 
compact systems become very important even for protons. 
\vspace{0.2cm}

{\it AGN cores and jets}~\\
Blast wave in AGN jets have typical sizes of a few percent of a parsec with magnetic fields of the 
order of 5 gauss.\cite{Zas} They could in principle  
lead to a maximum energy of a few tens of EeV. Similarily for AGN cores with a size of a few 
$10^{-5}$~pc and a field of order $10^3$~G one reaches a few tens of EeV.
However those maxima, already marginal, are unlikely 
to be achieved under realistic conditions. The very high radiation fields in and around the 
central engine of an AGN will interact with the accelerated protons 
producing pions and $e^+e^-$ pairs. Additional energy loss due to synchrotron radiation 
and Compton processes lead to a maximum energy of about \E{16}, much below 
the initial value.\cite{Batt_Sigl} To get around this problem, the 
acceleration site must be away from the active center and in a region with a lower radiation density such as in the terminal shock sites of the jets: a requirement possibly fulfilled by FR-II radio galaxies.
\vspace{0.2cm}

{\it FR-II radio galaxies}~\\
Radio-loud quasars are characterized by a very powerful central engine ejecting matter along 
thin extended jets. 
At the ends of those jets, the so-called hot spots, the relativistic shock wave is believed to be able to accelerate particles
up to ZeV energies. This estimate depends strongly on the value assumed for the spots' local
magnetic field, a very uncertain parameter. Nevertheless FR-II galaxies seem the best 
potential astrophysical source of UHECR.\cite{Biermann} Unfortunately, no nearby (less than 100~Mpc) 
object of this type is visible in the direction of the observed highest energy events. 
The closest FR-II source, actually in the direction of the Fly's Eye event at 320~EeV, is at about 
2.5~Gpc, way beyond the GZK distance cuts for nuclei, protons or photons.
\vspace{0.2cm}

{\it Gamma Ray Burst}~\\
Gamma ray bursters (GRB) are intense source of gamma rays of a few milliseconds with gamma energies
ranging from about 1 KeV to a few GeV. Several hundreds have been observed by satellites. 
The most favored GRB emision model is the ``expanding fireball model''
where one assumes that a large fireball, as it expends, becomes optically thin
hence emitting a sudden burst of gamma rays. The engine (the power source) of such a fireball 
remains unknown whilst the explanation of the non thermal spectra observed needs some 
additionnal modeling (such as internal shocks in the expanding fireball). 
\par
The observation of afterglow (low energy gamma ray emission of the heated gas in which the fireball
expanded) allowed to measure the red shift of the GRBs from which one confirmed their 
cosmological origin (and a support for the fireball model). Under certain conditions,
GRB can be shown to accelerate protons up to \E{20} therefore making them a good candidate site
for UHECR production. However in such a framework the UHECR spectrum should clearly
show the GZK cut-off while above \E{20} the distribution of arrival directions should be strongly 
anisotropic. Although more data is needed the 20 events already observed above \E{20} 
do not seem to confirm this hypothesis. In addition, the detection of high energy neutrinos 
(\E{14} and eventually \E{18} depending on the GRB environment) in coincidence with the gamma 
burst would be a strong evidence for this model\cite{Waxman}.  
\end{quote}

\subsubsection{Additional constraints}\noindent
In addition to the constraint given by Eq.~(\ref{eq:Hillas}), candidate sites must also 
satisfy two additional conditions.
\begin{itemize}
\item The acceleration must occur on a reasonable time scale, e.g. the size of the acceleration 
region must be less than the interaction length of the accelerated particle. This is a 
relatively weak constraint since all the objects in the Hillas 
diagram have a size below 1 Mpc. However, in shock acceleration mechanisms the rate 
of energy loss on the CMB must be less than the rate of energy gain: 
\begin{equation}
-\frac{dE_{\text{loss}}}{dt} \propto \text{const}\times E < \frac{dE_{\text{gain}}}{dt} 
\end{equation}

\item The acceleration region must be large enough so that synchrotron losses are negligible compared to the 
energy given by Eq.~(\ref{eq:Hillas}). For shock acceleration the radiated synchrotron power must be below the rate of 
energy gain:
\begin{equation}
-\frac{dE_{\text{sync}}}{dt} \propto \frac{E^4}{R^2}\propto  B^2~E^2
\end{equation}
\end{itemize}

\begin{figure}[!htb]
\vspace*{13pt}
\begin{center}  
\epsfig{file=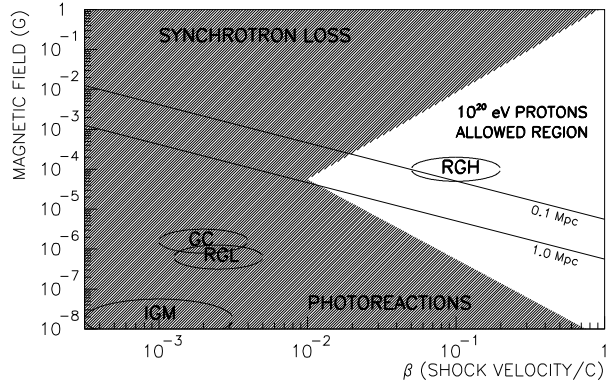, width = 8.5cm}
\end{center}
\vspace*{13pt}
\caption{Magnetic field strength and shock velocity of possible sites. 
GC refers to Galactic Cluster (accretion shocks), IGM to Inter 
Galactic Medium, RGL to Radio Galaxy Lobes and RGH to Radio Galaxy Hot Spots
(a subclass of RGL).\label{adrf23}}
\end{figure}

Using, as the characteristic acceleration time, $T_A = R / \beta$ (where $\beta$ is 
the shock velocity) one finds a characteristic gain rate of :
\begin{equation}
\frac{dE_{\text{gain}}}{dt} \approx \frac{E_{\text{max}}}{T_A} \propto \beta^2~B.
\end{equation}
For a given $E_{\text{max}}$ (e.g. 100~EeV), these two constraints define two lines in the 
$\log{B}$, $\log{\beta}$ plane above and below which particles cannot be accelerated at 
the required energy. As can be seen on Figure~\ref{adrf23} the only
remaining candidates are the radio galaxy hot spots (RGH).

\vspace{0.5cm}
\par
\noindent
To conclude on the bottom-up scenario, let us mention a recent analysis from Farrar and 
Biermann.\cite{Farrar_Biermann} They have shown that, on cosmological scales, the 
correlation between the arrival direction of the five highest energy events and 
Compact Quasi Stellar Objects (CQSO's) which include radio-loud galaxies is 
unlikely to be accidental. However, only a new type of neutral particle could travel 
on distances over 1~Gpc without 
losing its energy on the CMB nor being deflected by extragalactic magnetic fields. 
More data at very high energy are needed to validate this result which would sign the existence 
of a new particle physics phenomenon. 
\newcommand{\X}{$X$}
\subsection{``Exotic" sources: Top-Down scenario}\label{TopDown}\noindent
One way to overcome the many problems related to the acceleration of UHECR, 
their flux, the visibility of their sources and so on, is to 
introduce a new unstable or meta-stable super-massive particle, currently called the \X-particle. 
The decay of the \X-particle produces, among other things, quarks and leptons. 
The quarks hadronize, producing jets of hadrons which, together with the decay products of the 
unstable leptons, result in a large cascade of energetic photons, neutrinos and light leptons with a small 
fraction of protons and neutrons, part of which become the UHECR. 

For this scenario to be observable three conditions must be met:
\begin{itemize}
\item The decay must have occurred recently since the decay products must have traveled less than about 100~Mpc because of the attenuation processes discussed above.
\item The mass of this new particle must be well above the observed highest energy (100~EeV range), 
a hypothesis well satisfied by Grand Unification Theories (GUT) whose scale is around $10^{24}$-\E{25}.
\item The ratio of the volume density of this particle to its decay time must be 
compatible with the observed flux of UHECR. 
\end{itemize}
\noindent
The \X-particles may be produced by way of two distinct mechanisms:
\begin{itemize}
\item  Radiation, interaction or collapse of Topological Defects (TD), producing \X-particles that
 decay instantly. In those models the TD are leftovers
from the GUT symmetry breaking phase transition in the very early universe.
Quantitative predictions of the TD density that survives a possible inflationary 
phase rely on a large number of theoretical hypotheses. Therefore they cannot be taken as face value,
although the experimental observation of large differences could certainly be interpreted as 
the signature of new effects.

\item  Super-massive metastable relic particles from some primordial quantum field, produced after
the now commonly accepted inflationary stage of our Universe. Howerver the ratio of their lifetime 
to the age of the universe requires a fine tuning ($10^{-11}$) with their relative abundance 
as is discussed in section\ref{moreAboutX}.
It is worth noting that in some of those scenarios the relic particles may also act as non-thermal 
Dark Matter.
\end{itemize}

In the first case the \X-particles instantly decay and the flux of UHECR is related to their production rate
given by the density of TD and their radiation, collapse or interaction rate, 
while in the second case the flux is driven by the ratio of the density of the relics over their 
lifetime. In the following
the  terms  {\it ``production or decay rate''} will refer to these two situations.
Before discussing the exact nature of the \X-particles we shall briefly review the main characteristics
of the decay chain and the expected flux of the energetic outgoing particles.
\subsubsection{\X decay and secondary fluxes}\noindent
At GUT energies and if they exist,  squark and 
sleptons are believed to behave like their corresponding super-symmetric partners 
so that the gross characteristics of the cascade may be inferred from the 
known evolution of the quarks and leptons. Of course the internal mechanisms of the decay and 
the detailed dynamics of the first secondaries do depend on the exact nature 
of the particles but the bulk flow of outgoing particles is most certainly independent of such details.\cite{Batt_Sigl} 

\begin{figure}[!htb]
\vspace*{13pt}
\begin{center}  
\epsfig{file=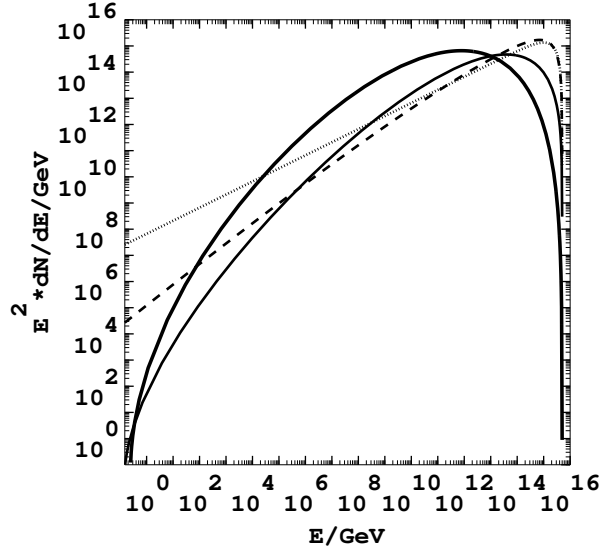, width = 8cm}
\end{center}
\vspace*{-13pt}
\caption{Fragmentation function in the Modified Leading Log Approximation for a total 
jet energy of $5\times$\E{24} with SUSY (thick solid line peaking at $10^{12}$GeV) and 
without SUSY (thin solid Line) as calculated by Bhattacharjee and Sigl. 
Other lines are Hill's formula (dashed) and an approximated expression (dotted).  
Note that around \E{20}, our region of interest, the fluxes are not too widely different.
\label{Sigl23}}
\end{figure}

A common picture for the  hadronisation of the decay products follows three steps. 
At the high energy end, the perturbative QCD-inspired recipes provide a good framework for the description 
of the hard processes driving the dynamics of the parton cascade. At a cutoff energy 
of about 1~GeV soft processes become dominant and partons are glued together to form 
color singlets which will in turn decay into known hadrons. The LUND\cite{Lund}
string fragmentation model provides a description for the second and last phases while 
a model like the Local Parton-Hadron Duality directly relates the parton density in the parton cascade to 
the final hadron density.\cite{Lphd} Nevertheless and despite the fact that up to 40\% 
of the initial energy may turn into LSP, the cascade produces a rather 
hard\footnote{For a power law spectrum of exponent $\alpha<2$ the total energy ($\propto E^{2-\alpha}$)
is dominated by the high energy end of the integral, i.e. a few very energetic particles, thus a hard spectrum, while for $\alpha>2$ the energy is carried by the very large number of low energy particles, i.e. a soft spectrum.}
~hadron spectrum adequately described by:
$$ \frac{dN_h}{dE} \propto E^{-\alpha}\mbox{~~with~~} 1<\alpha<2$$
in the range $E/m_X \ll 1$, where $m_X$ is the \X-particle mass. At the high energy end a cutoff occurs 
at a value depending on the \X-particle 
mass and on the eventual existence of new physics such as Super Symmetry (SUSY), which would displace 
the maximum of the hadron spectrum to a lower energy (see Figure~\ref{Sigl23}).

Indeed, Super Symmetry is not the only candidate theory for new physics beyond the standard 
model, although the only known acceptable one. Other (yet unknown) models may appear 
as possible alternatives in the future.  However, in all cases, secondaries from Top-Down mechanisms should
manifest themselves as a change of slope in the UHECR spectrum, above 10~EeV and over a range which 
will reflect the (new) physics at play. 

In all conceivable Top-Down scenarios, photons and neutrinos dominate at the end of the hadronic cascade. 
This is \emph{the} important distinction from the conventional acceleration mechanisms.
The spectra of  photons and neutrinos can be derived from the charged 
and neutral pion densities in the jets as:
$$\Phi_{\gamma}^{\pi^0}(E,t) \simeq 2 \int_E^{E_{\text{jet}}}\Phi_{\pi^0}(\varepsilon,t)d\varepsilon/\varepsilon$$
$$\Phi_{\nu}^{\pi^\pm}(E,t) \simeq 2.34 \int_{2.34E}^{E_{\text{jet}}}\Phi_{\pi^\pm}(\varepsilon,t)d\varepsilon
/\varepsilon$$
where $E_{\text{jet}}$ is the total energy of the jet (or equivalently the initial parton energy). Since 
$\Phi_{\pi^\pm}(\varepsilon,t)\simeq 2\Phi_{\pi^0}(\varepsilon,t)$, photons and neutrinos should have very 
similar spectra. These injection spectra must then be convoluted with the transport phenomena to obtain 
the corresponding flux on Earth. As was mentioned in Section \ref{photon} the photon 
transport equation strongly depends on its energy and on the badly known Universal 
Radio Background and extragalactic magnetic fields. 

\subsubsection{$X$ production or decay rates: a lower limit}\noindent
The production or decay rates of the \X-particles are very model dependent and no firm  
prediction on the expected flux of UHECR can be made. However, in their review, Bhattacharjee 
and Sigl evaluate with a simple model the rate needed to explain the observed UHECR 
fluxes. Assuming that photons dominate at the source and on Earth and that they follow 
a power law spectrum of index $\alpha$; assuming also that the 
initial \X~decay secondaries are quarks and leptons in equal numbers, they calculate 
a lower limit on the production rate given by (for $\alpha=1.5$):
\begin{equation}
\dot{n}_X \geq 10^{-46} \left( \frac{\mbox{10\,Mpc}}{l_E(E_\gamma)}\right )\left( \frac{E^2j_\gamma(E
)}{{\cal F}_\oplus}\right )\sqrt{\frac{10^{16}\text{GeV}}{m_X}} ~\text{cm}^{-3}\text{s}^{-1}
\label{prod_lim}
\end{equation}
Here ${\cal F}_\oplus \approx 1$~eV$\,$cm$^{-2}$s$^{-1}$sr$^{-1}$ is the observed 
energy flow of UHECR at 100~EeV and $l_E(E_\gamma)$ the photon attenuation length. 
Additional normalization factors of order unity have not been reproduced here. 
In other words, for TD or relics to explain the observed UHECR
flux at 100~EeV and assuming an \X~mass of $10^{16}$~GeV their production or decay rates 
must be larger than $10^{-46} \text{cm}^{-3}\text{s}^{-1}$. This is of course only 
an order of magnitude calculation which may be modified by 
the decay dynamics and the distribution of the \X-particles, but can be used 
as a reasonable scale of the necessary rates.
 
\subsubsection{More about \X-particles}\noindent\label{moreAboutX}
{\it Topological defects}\\
The very wide variety of topological defect models together with their large number of parameters makes them 
difficult to review in detail. Many authors have addressed this field. Among them, let us mention 
Vilenkin and Shellard\cite{Shell} and Vachaspati\cite{Vach1,Vach2} for a review on TD formation 
and interaction, and Bhattacharjee,\cite{Bhatt} Bhattacharjee and Sigl\cite{Batt_Sigl} 
and Berezinky, Blasi and Vilenkin\cite{Berez} for a review on experimental signatures in 
the framework of the UHECR. 

According to the current picture on the evolution of the Universe, several symmetry 
breaking phase transitions such as $GUT\Longrightarrow H~...\Longrightarrow SU(3)\times 
SU(2)\times U(1)$ occurred during the cooling. For those ``spontaneous'' symmetry 
breakings to occur, some scalar field (called the Higgs field) must acquire a non vanishing 
expectation value in the new vacuum (ground) state. Quanta associated to those fields have 
energies of the order of the symmetry breaking scale, e.g. $10^{15}-10^{16}$~GeV for the 
Grand Unification scale. Such values are indeed perfectly in the range of interest 
for the above mentioned \X-particles.

During the phase transition process, non causaly connected regions may evolve towards different 
states - the correlation length is smaller than the horizon - in such a way that at the different 
domain borders, the Higgs field is forced to keep a vanishing expectation value for topological 
reasons. Energy is thus trapped at the border called a TD whose properties depend on the topology of 
the manifold where the Higgs potential reaches its minimum (the vacuum manifold topology). 

Possible TDs are classified according to their dimensions: magnetic monopoles (0-dimensional, point-like); 
cosmic strings (1-dimensional); a sub-variety of the previous which carries current and is superconducting; 
domain walls (2-dimensional); textures (3-dimensional). Among those, only monopoles 
and cosmic strings are of interest as possible UHECR sources: textures do not 
trap energy while domain walls, if they were formed at a scale that could explain EHECR, 
would over-close the Universe.\cite{Zel}

In GUT theories, magnetic monopoles always exist because the reduced symmetry group contains at least the 
electroweak $U_Y(1)$ invariance. In fact it is the predicted overabundance of magnetic monopoles 
in our present universe that led Guth\cite{Guth} to come up with the now well adopted idea of an 
inflationary universe. Strings on the other hand are the only defects that can be relevant for structure 
formation. It is possible, from the scaling property of the string network,
to relate the string formation scale $\eta$ to the mass fluctuations in the Universe.
Using the large scale mass fluctuation value of $\delta M/M \sim 1$ this 
gives $\eta\simeq10^{16}$~GeV and similar  conclusions are drawn if one uses the COBE results 
on CMB anisotropies.\cite{Brand} 
It is striking to see that if strings were to play a role in large scale structure formation, 
hence making the Hot Dark Matter scenario viable, 
\bi
\item the proper energy scale is precisely the grand unification scale of GUT theories,
\item this scale also corresponds to the one relevant for UHECR production.
\ei
\noindent
When two strings intercommute, the energy release sometimes leads to the production of small 
loops that will release more energy when they collapse. These are, among other mechanisms,
fundamental dissipation processes that prevent the string network from dominating the energy 
density in the Universe. For monopoles, it is the annihilation of monopolonia (monopole-antimonopole bound 
states)\cite{Hill,Schramm_Hill}  that releases energy\footnote{In fact monopolonia  
are too short lived but monopole-anti-monopole pairs connected by a string have appropriate lifetime. 
This happens when the $U(1)$ 
symmetry  is further broken into $Z_2$} -although the existence of monopole of the proper energy scale 
is very questionable as they are either over abundant or washed out by inflation-. 
In each case part of the released  energy is in the form of \X-particles.    

Strings and monopoles come in various forms according to the scale at which TDs are formed
and to the vacuum topology. They may even coexist. Nevertheless, the \X-particle production rate may, 
on dimensional grounds, be parameterized in a very general way.\cite{Bhatt_Hill_Schramm} 
Introducing the Hubble time $t$, the production rate can be written as:
\begin{equation}
\dot{n}_X(t) = \frac{Q_0}{m_X}\left (\frac{t}{t_0}\right )^{-4+p}
\label{X_prod}
\end{equation}
where $Q_0\equiv \dot{n}_X(t_0)\,m_X$ is the energy injection rate at $t=t_0$ (the present epoch). The parameter $p$ 
depends on the exact TD model. In most cases (intercommuting strings, collapsing loops as well as 
monopolonium annihilation) $p=1$ but superconducting string models can have $p\leq 0$ while decaying 
vortons\footnote{Superconducting string loops stabilized by the angular momentum of the charge 
carriers.}~~give $p=2$.  

One can compare the integrated energy release of Eq.~(\ref{X_prod}) in the form of 
low energy (10~MeV - 100~GeV) photons resulting from the cascading of the electromagnetic component 
of the $X$-particle decay with the diffuse extragalactic gamma ray 
background, $w_{\text{em}}\sim 10^{-6}$~eV~cm$^{-3}$~s$^{-1}$, as measured by EGRET. 
Assuming as in Ref.\cite{Berez} that half of the energy release goes into the electromagnetic 
component, one obtains~:
\begin{equation}
w_{\text{em}} = \frac{Q_0}{2}\int_{t_{\text{min}}}^{t_0} \left( \frac{t}{t_0}\right )^{-4+p} \frac{dt}{(1+z)^4}.
\label{wem_prod}
\end{equation}
where $(1+z) = (t_0/t)^{2/3}$ in a matter dominated Universe.
For $\alpha\equiv p -1/3 > 0$ evolutionary effects are negligible and Eq.~(\ref{wem_prod}) simply leads to 
$$
w_{\text{em}} \simeq \frac{Q_0}{2}\frac{t_0}{\alpha}
\label{wem_1}
$$
or, using the EGRET limit and $t_0\simeq 2\times 10^{17}h^{-1}$s, to:
$$
\dot{n}_X(t_0) \leq \alpha \frac{10^{-48}h~\text{cm}^{-3}\text{s}^{-1}}{m_X/10^{16}\,\text{GeV}}
$$
a limit hardly compatible with the order of magnitude given by 
Eq.~(\ref{prod_lim}). However, more information about the UHECR fluxes, 
the diffuse gamma ray background and extragalactic magnetic fields are needed to confirm this.

In the models where $\alpha<0$ evolutionary effects can become important. In fact it is the lower bound of the 
integral of Eq.~(\ref{wem_prod}) that would dominate. Using as a lower bound the decoupling time
$t_{\text{dec}}/t_0\sim 10^{-5}$ one gets:
$$
w_{\text{em}} \simeq \frac{Q_0}{2}\left (\frac{t_{\text{dec}}}{t_0}\right )^{-\alpha}\frac{t_0}{|\alpha|}
\label{wem_2}
$$
or,  
$$
\dot{n}_X(t_0) \leq |\alpha|\frac{10^{-48-5\alpha}\text{cm}^{-3}\text{s}^{-1}}{m_X/10^{16}\,\text{GeV}}.
$$
which,  for $p\leq 0$ is perfectly compatible with Eq.~(\ref{prod_lim}). 
However the large density of gamma rays 
released in the early Universe impacts on the $^4$He production and on the uniformity of the CMB making this 
kind of models currently unfavored in the context of UHECR.

\vspace*{0.5cm}
\noindent
{\it Supermassive relics}\\
Supermassive relic particles may be another possible source of UHECR.\cite{Berez1} Their mass should be 
larger than $10^{12}$~GeV and their lifetime of the order of the age of the Universe since 
these relics must decay now (close by) 
in order to explain the UHECR flux. Unlike strings and monopoles, but like monopolonia, 
relics aggregate under the effect of gravity like ordinary matter and act as a (non thermal) 
cold dark matter component. The distribution of such relics should consequently be biased 
towards galaxies and galaxy clusters. A high statistics study of the UHECR arrival distributions 
will be a very powerful tool to distinguish between aggregating and non-aggregating Top-Down sources. 

If one neglects the cosmological effects, a reasonable assumption on the decay rate would 
simply be, since the decay should occur over the last 100 Mpc/c:
$$
\dot{n}_X = \frac{n_X}{\tau}
$$
where $\tau$ is the relic's lifetime and where the relic density $n_X$ may be given in 
terms of the critical density of the Universe $\rho_c$ as:
$$
n_X = \frac{\rho_c(\Omega_X h^2)}{m_X} = 10^{-17} (\Omega_X h^2) \left (\frac{m_X}{10^{12}~\text{GeV}}\right )^{-1}
$$   
From which, with the constraint given by Eq.~(\ref{prod_lim}) and using $m_X=10^{12}$~GeV, one obtains a 
lifetime of the order of 
$10^{21}(\Omega_X h^2)$ years. To obtain such a value, orders of magnitude larger than 
the age of the Universe, one needs a  symmetry (such as $R$-parity) to be very softly 
broken unless the  fractional abundance 
$\Omega_X$ represents only a tiny part ($\sim 10^{-11}$) of the density of the Universe, in which case the 
 production mechanism of relics must be extraordinarily inefficient.
\subsection{Conclusions}\noindent
The cosmic rays' chemical composition, the shape of their energy spectrum and the distribution of 
their directions of arrival will prove to be powerful 
tools to distinguish between the different acceleration or decay scenarios. 

If the UHECR are conventional hadrons accelerated by Bottom-Up mechanisms, they should point 
back to their sources, with a quite specific distribution in the sky and a spectrum clearly 
showing the GZK cutoff.  If, on the other hand,  the accelerated particles 
are not conventional, they should be neutral particles in order not to interact with the CMB 
(they can only be secondary collision products therefore putting even more requirements on the source power) 
and must interact strongly with the atmosphere. 

For Top-Down mechanisms and above the ZeV, one should observe a flux of photons 
(and neutrinos) as the photon absorption length increases (up to several Gpc). Below 
100~EeV the spectrum shape will depend on the relative values of, the characteristic distance between 
TD interactions or relic particle decays and Earth ($D$), the proton
attenuation length ($R_p$), and the photon absorption 
length ($L_\gamma$).  
Following the description of Ref.\cite{Berez} the following situations can be disentangled:
\begin{itemize}
\item $R_p<D$: a very low flux with an exponential cutoff. If the sources are nearby,
the observed distribution will be strongly anisotropic.

\item$L_\gamma<D<R_p$: the protons dominate and the GZK cutoff is visible. As energy 
increases, the direction of arrival distribution
 should become more and more anisotropic as photons no longer get absorbed.

\item $D<L_\gamma$: a very strong flux in the direction of the sources; photons dominate.

\item $D\ll L_\gamma$: the GZK cutoff is visible and protons dominate as long as $R_p(E)$ is much 
larger than $L_\gamma(E)$. Photons dominate  above a few ZeV. The arrival distribution is isotropic 
at all energies.
\end{itemize} 
For relic particles and TDs like vortons and monopolonia, because of the accumulation in the galactic 
halo, photons will dominate the flux. Some anisotropy 
should be visible due to the earth's  slightly eccentric position in the halo. The spectrum will not 
show any GZK cutoff and the EGRET constraints on the injection 
rate will not apply as the emitted photons have no time to cascade over the short distances.

Finally, if nuclei can possibly be UHECR candidates in Bottom-Up scenarios, they are completely 
excluded in the Top-Down cascades. 

\section{Extensive Air Showers Phenomenology\label{ident}}\noindent
Extensive Air Showers (EAS) are the particle cascade following the interaction 
of a cosmic ray particle with an atom of the upper atmosphere.
On an incident cosmic ray the atmosphere acts as a calorimeter with
variable density, a vertical thickness of 26 radiation lengths and
11 interaction lengths. In the following, we will describe the
properties of a vertical EAS initiated by a 10~EeV proton and mention how
some of these properties are modified with energy and with the nature of
the initial cosmic ray.

\subsection{Shower development, size and particle content}
\begin{figure}[!htb] 
\begin{center}
\epsfxsize=25pc
\epsfbox{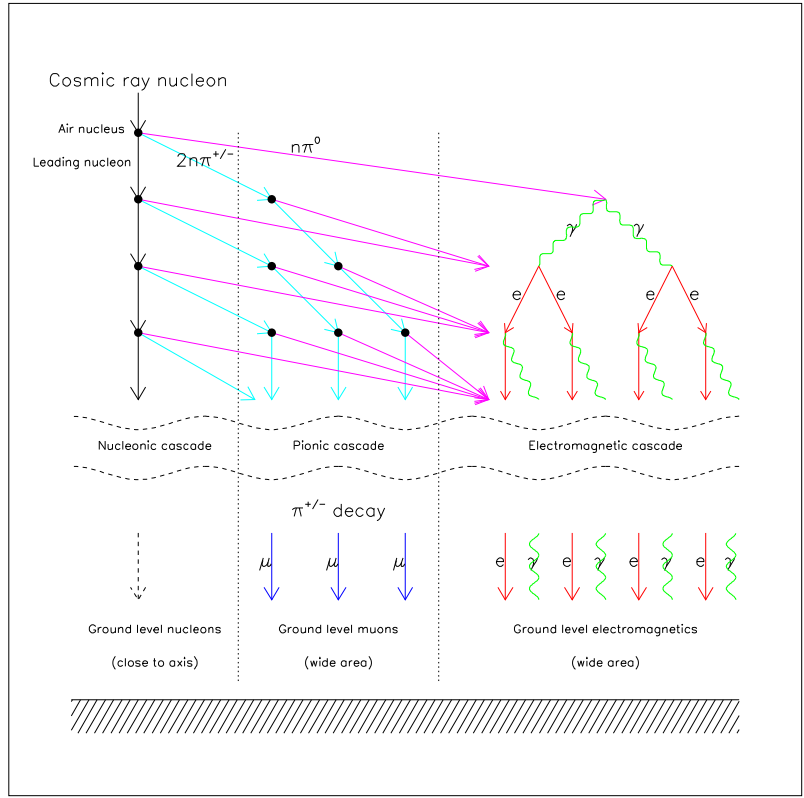} 
\end{center}
\caption{Schematic development of an atmospheric shower. Three components are 
depicted (case of an hadronic primary).Hadronic cascade (leading nucleon) close to the axis (100m), 
EM cascade ( $\pi^0$ decay) and pion cascade,the last 2 component extend a few km from the axis.
\label{schematics}}
\end{figure}

\noindent
A schematic development of an atmospheric shower is shown on Figure~\ref{schematics}. 
At sea level (atmospheric thickness of 1033~g/cm$^2$) the number of
secondaries reaching ground level (with energies in excess of 200~keV) is about $3\times 10^{10}$
particles. 99\% of these are photons and electrons/positrons in a ratio
of 6 to 1. Their energy is mostly in the range 1 to 10~MeV and they
transport 85\% of the total energy. The remaining 1\% is shared between
mostly muons with an average energy of 1~GeV (and carrying about 10\% of
the total energy), pions of a few GeV (about 4\% of the total energy)
and, in smaller proportions, neutrinos and baryons. The shower footprint (more than 1 muon
per $m^2$) on the ground extends over a few $km^2$

At each step of the cascade the hadronic energy is shared between 70\% hadronic and 
30\% electromagnetic. The shower grows until the average pion energy reaches a critical value 
($E_c\sim GeV$) where their interaction length is longer than their decay length. At this stage the 
shower development is at a maximum (about 830$g/cm^2$ of atmospheric depth) and starts to 
very slowly decrease. 
\par
\begin{figure}[!htb] 
\vspace*{0.5cm}
\begin{center} 
\epsfxsize=20pc
\epsfbox{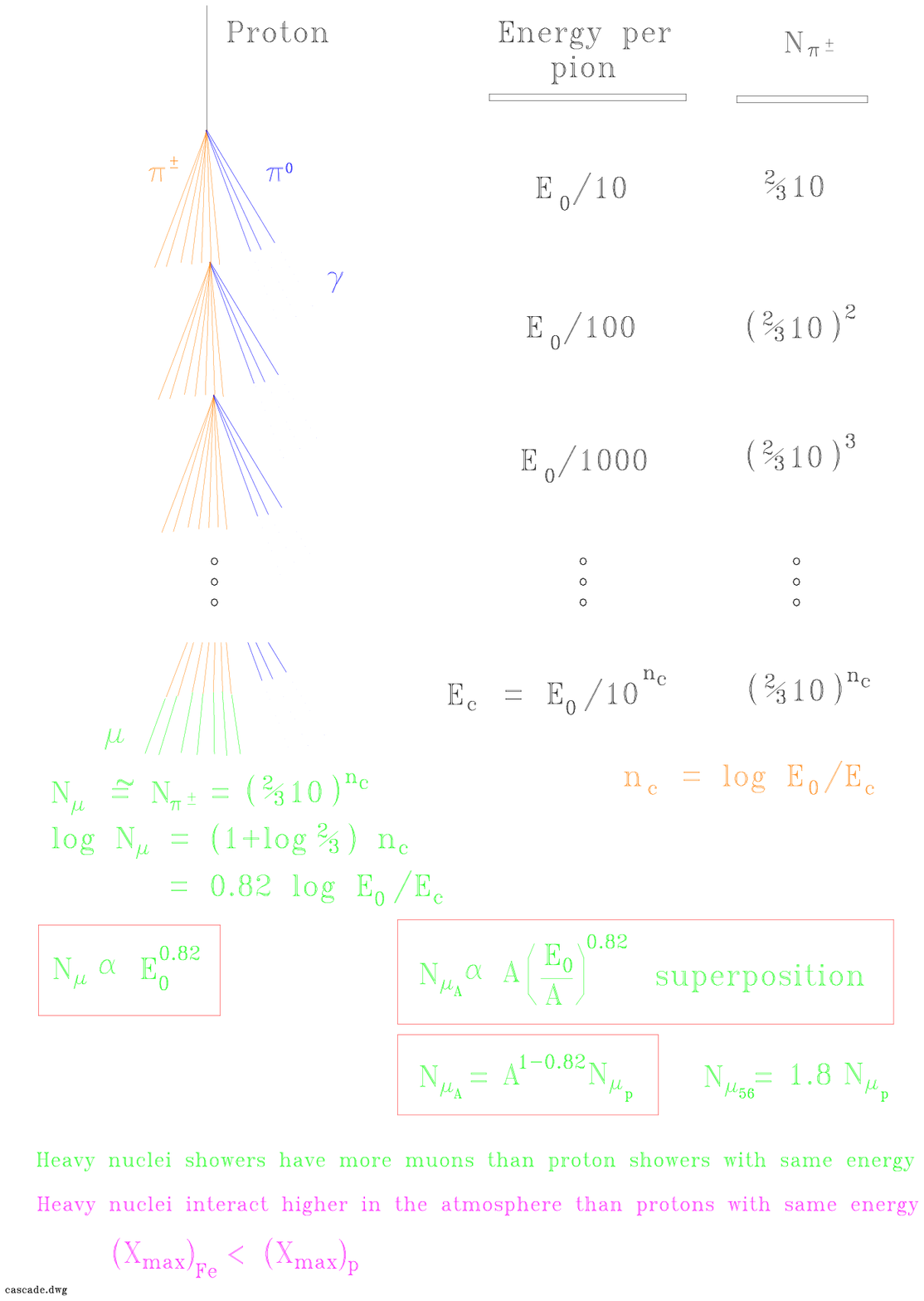}
\end{center}
\caption{A simplified shower development model.\label{simpleModel}}
\end{figure}

Using a very simplified model, as depicted on Figure~\ref{simpleModel} 
one can derive the main 
EAS properties. Let $\lambda~$ be the interaction length in Air,
$E_c~$ the critical energy below which particles only decay or loose energy via ionization and 
$X$ the shower depth in the atmosphere (usually measured in $g/cm^2$) then~:
\bi
\item the particle number ($N(X)$) at a given depth $X$ grows like $N_s^{X/\lambda}$  (where
$N_s$ is the number of secondaries in an interaction),
\item the secondaries mean energy $E(X)$ at depth $X$ is $E_0/N(X)$
\item and since by definition $E_c = E_0/N(X_{max}) = E_0/N_{max}$ one can show that the 
position of the shower maximum $X_{max}$ varies like $\lambda~log(E_0)$ while the size 
at maximum $N_{max}$ is proportionnal to the primary energy.
\ei  
\vspace*{0.5cm}

At ground level most of the energy is caried out by photon and electrons and their number 
is proportionnal  to the total shower energy, i.e. to the primary cosmic ray energy.
$$ E_{EM} \propto (N_\gamma + N_{e^+e^-})<\epsilon> \quad \propto E_{prim}$$ 
where $<\epsilon>$ is the mean photon or electron/positron energy at ground level ($\sim 10MeV$).

\par 
The muon content of the shower does not scale linearly with energy. Muons are mainly
produced through pion decay whose energy increases faster ($\propto E_{prim}$) than their number 
($\propto \log{E_{prim}}$) therefore the number of pions reaching the critical energy and decaying into 
muons is not proportionnal to the primary energy. With a simple Monte Carlo one can show that for 
a proton primary the muon number scale as~: 
$$ N_\mu^p(E) \propto E^{0.85}.$$
\par
For nuclei, the {\it superposition principle} stipulates that a nucleus $^AN$ of total energy $E_0$ is
equivalent to $A$ proton of energy $E_0/A$. Therefore the muon content of an EAS initiated 
by a nucleus will contain more muons than the EAS initiated by a single proton of the same energy~:
$$ N_\mu^A(E) \propto A^{0.15}\times N_\mu^p(E).$$
Therefore an Iron primary ($^{56}Fe$) gives 80\% more muons then a proton of the same energy.  
\par 
For lighter primaries such as photons, the muon component will be much smaller as the number of 
pion produced in the cascade is greatly reduced. The position of the shower maximum will also 
strongly depends on the primary type (photon or neutrinos) and on additionnal phenomena such as 
conversion in the geomagnetic field and the Landau Pomeranchuk Migdal effect\cite{LPM}. 
This effect describes the decrease of the photon/electron nucleus cross-sections with energy
and with the density of the medium with which they interact.   

\par
Let's conclude by stressing that the experimental measurements of both the EAS muon content and
maximum depth in the atmosphere are of the outmost importance to derive informations about the 
primary cosmic ray nature.

\subsection{Spatial structure}\noindent
An EAS is essentially a thin disk (a few $\mu s$ thickness) of particles moving 
at the speed of light. The longitudinal and the lateral development as well as the time 
structure of the shower are characteristics of its nature. In the following we'll 
describe the dominant features of those profiles.
\par
As was previously mentionned, the longitudinal development is characterized
by a maximum reached at an atmospheric depth of 830 g/cm$^2$ (or an equivalent 
altitude of about 1800 meters for vertical showers). At maximum, the shower contains 
about $7\times 10^9$ electrons. The depth of maximum is a function of the primary energy \underline{and}
type,  
$$X_{max}(10\times E_0) = X_{max}(E_0) + 55~(g/cm^2)$$
and, from the superposition principle,
$$X_{max}(^{56}F_e;E_0) = X_{max}(p;E_0) - 100~(g/cm^2) $$

\begin{figure}[!htb] 
\begin{center} 
\epsfxsize=20pc
\epsfbox{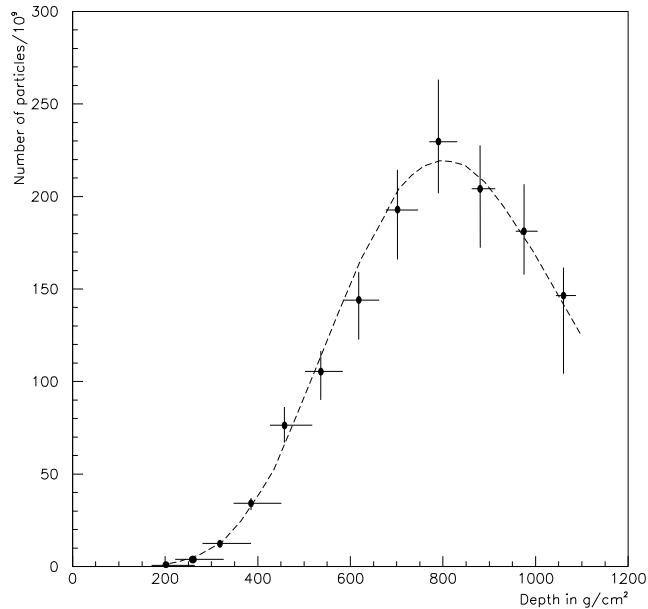}
\end{center}
\caption{A longitudinal profile as measure by the Fly's Eye experiment. The measured energy is
$3\times$\E{20}.\label{longitudinal}}
\end{figure}

The electrons contained in the shower excite the Nitrogen molecules of
the atmosphere which produce fluorescence light. As the emision of light is 
proportionnal to the number of ionizing particles, the fluorescence light follows the 
longitudinal developmewnt of the shower. Such a profile as measured by the Fly's Eye 
optical device is shown on Figure~\ref{longitudinal}.

\par
The lateral development of the shower is represented by its Moli\`ere radius
(or the distance within which 90\% of the total energy of the shower is
contained) which, in ``standard air'' is 70~m. However, the actual
extension of the shower at ground level is of course much larger. As an
example, at a distance of 1~km from the shower axis, the average
densities of photons/electrons/muons are 30/2/1 per m$^2$ respectively. 
\par
\begin{figure}[!htb] 
\begin{center} 
\epsfxsize=25pc
\epsfbox{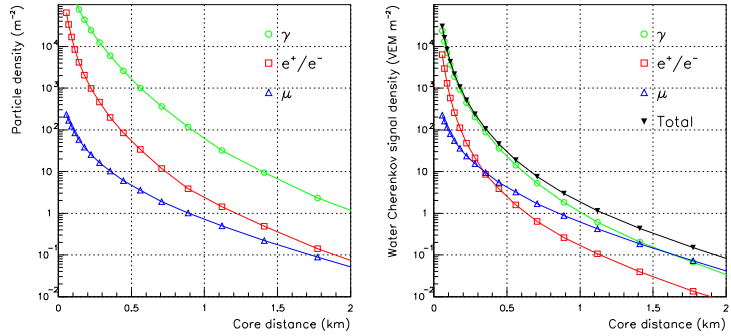}
\end{center}
\caption{Particle densities as a function of core distance (left). Corresponding detector response 
for a water Cherenkov tank (right).\label{densities}}
\end{figure}

The particle density as a function of distance to the shower axis is parameterized by the ``lateral distribution function''~:
$$ \rho(r) \propto k\times r^{-[\eta+f(r)]}$$ 
where $f$ and $k$ depend on the ground detector type and $\eta$ upon the primary incident angle and energy. For $r>800m$ this (empirical) expression must be modified by a factor $(r/800)^{1.03}$ to better 
modelize the experimental data. Patricle densities are presented on figure~\ref{densities} as well
as the corresponding response of water Cherenkov detector.

\par
The particle arrival time at a given detector is essentially driven 
by geometrical arguments. Defining the shower plane as the plane tangent to the shower front 
and perpendicular to its axis one observes that far from the core particles arrive after
the shower plane. In fact the shape of the shower front above the shower plane can be 
modeled with a cone. 
\bi
\item At a distance R from the core particles are spread over a time interval which is 
roughly proportionnal to R. Moreover, this time spread increases when $X_{max}$ increases. 
\item Muons, as they can travel quite far and almost straight, arrive 
in general earlier than the electromagnetic component.
\ei
The first item allows to distinguish small close-by showers from far away large ones while the second 
allows to count muons without a particle identification detector. As an example a $^{56}Fe$ primary 
gives more muons and interacts sonner (higher) in the atmosphere. Consequently, the signal rise time
(for a signal proportionnal to the particle number) will be slower than 
for a proton shower.
\par
Let $t_\alpha$ be the time for which $E(t_\alpha) = \alpha \times E(+\infty)$ with $\alpha \in [0,1].$
$E(t)$ is the integrated detector signal amplitude as a function of time. Then
$ t_{\alpha-\beta} = t_\beta-t_\alpha$ is an indirect measure of $X_{max}$ i.e. of the primary type.\\
  
\par\noindent{\bf Fluctuations~:}\\
All the above distributions are subject to fluctuations in the ground particle distibutions. Those 
fluctuations depend only weakly on the depth and characteristics of the first interaction. 
There is an optimal distance from the shower core 
where one can estimate the primary energy from the ground particle density. 
The shower to shower ``physical'' fluctuations due to the variation in the first interaction 
important near the core decrease with distance, while the statistical fluctuation in the densities
increase.

\subsection{Monte Carlo descriptions}
All these effects are studied through heavy use of EAS Monte Carlo programs
such as AIRES,\cite{Aires} CORSIKA,\cite{Corsika} HEMAS\cite{Hemas} or
MOCCA.\cite{Mocca} At the UHECR ranges, where the center-of-mass energies are 
much higher (almost two orders of magnitude) than those attainable in the future 
accelerator (like LHC),  the correct modelling of the EAS in these programs becomes delicate.
\par 
Some data are available from accelerator experiments such as HERA,\cite{Hera}
and showers of about $10^{16}$~eV are now being well studied through
experiments such as KASCADE.\cite{Kascade} The models are thus constrained at
lower energies and then extrapolated at higher ones.

The most commonly used models for the high energy hadronic interactions are
SIBYLL,\cite{Sibyll} VENUS,\cite{Venus} QGSJet\cite{QGSJet} and
DPMJet.\cite{DPMJet}  Interactions at lower energies are either processed
through internal routines of the EAS simulation programs or by well-known packages
such as GHEISHA.\cite{Gheisha}  Some detailed studies of the different models
are available.\cite{Comparison} The main shower parameters, such as the reconstructed 
direction and energy of the
primary CR, are never strongly dependent on the chosen model. However,
the identification of the primary is more problematic.
Whatever technique is chosen, the
parameters used to identify the primary cosmic ray undergo large physical
fluctuations which make an unambiguous identification difficult.

A complete analysis done by the KASCADE group on the hadronic core of
EAS\cite{Kascade} has put some constraints on interaction models beyond accelerator energies.
Various studies seem to indicate QGSJet as being the model which best
reproduces the data\cite{Comparison,Erlykin} with still some disagreement
at the {\em knee} energies ($10^{16}$~eV). For the highest energies,
additionnal work (and data) is needed to improve the agreement between the available models.

\section{Detection techniques}
\noindent
When the cosmic ray flux becomes smaller than 1 particle per m$^2$
per year, satellite borne detectors are not appropriate any more. This
happens above $10^{16}$ eV (the so-called ``knee" region). Then
large surfaces are needed and the detectors become ground-based. What they
detect is not the incident particle itself but the Extensive Air shower 
described in the previous section.
\par
All experiment aim to measure as accuratly as possible the three following quantities ~:
\bi
\item The primary direction (given by the shower axis),
\item the primary energy,
\item the primary nature (or mass).
\ei

\par
There are two major techniques used. The first, and the most frequent, is to build an array of sensors
(scintillators, water Cerenkov tanks, muon detectors) spread over a large area.
The detectors count the particle densities sampling
the EAS particles hitting the ground. The surface of the array
is chosen in adequation with the incident flux and
the energy range one wants to explore. From the timed sampling of the
lateral development of the shower at a given
atmospheric depth one can deduce the direction, the energy and possibly the identity
of the primary CR.
The second technique, until recently the exclusivity of a group from the University of
Utah, consists in studying the longitudinal development of the EAS by detecting
the fluorescence light produced by the interactions of the  charged secondaries.
                                                               
\subsection{The optical fluorescence technique}
\noindent
The basic principle is simple\cite{Cassiday} the fluorescence light which
is quickly and isotropically emitted by the nitrogen atoms of the atmosphere
can be detected by a photo-multiplier. 
The emission efficiency (ratio of the energy emitted as fluorescence 
light to the deposited one) is poor, less than 1\%, 
therefore observations can only be done on clear
moonless nights (which results in an average 10\% duty cycle) and low
energy showers can hardly be observed. However, at higher
energies, the huge number of particles in the shower\footnote{The highest
energy shower ever detected (320~EeV) was observed by 
the Fly's Eye detector: at the shower maximum, the number of particles was larger than
$2\times10^{11}$.}~ produce enough light to be detected even at large distances. 
\par 
The fluorescence yield is 4 photons per electron per meter at ground level pressure. The emitted
light is typically in the 300-400 nm UV range to which the atmosphere is quite
transparent. Under favorable atmospheric conditions an UHECR shower
can be detected at distances as large as 20 km, about two attenuation lengths 
in a standard desert atmosphere 
at ground level. 

The first successful detectors based on these ideas were built by a group of the University of Utah, 
under the  name of ``Fly's Eyes"\cite{Bird}, and used with the Volcano Ranch ground array\cite{Linsley}. 
A complete detector was then installed at Dugway (Utah) and started to take data in 1982. An updated
version, the High-Resolution Fly's Eye, or HiRes\cite{Hires}, is presently running on this same site. 

\begin{figure}[!htb] 
\begin{center}
\epsfxsize=15pc
\epsfbox{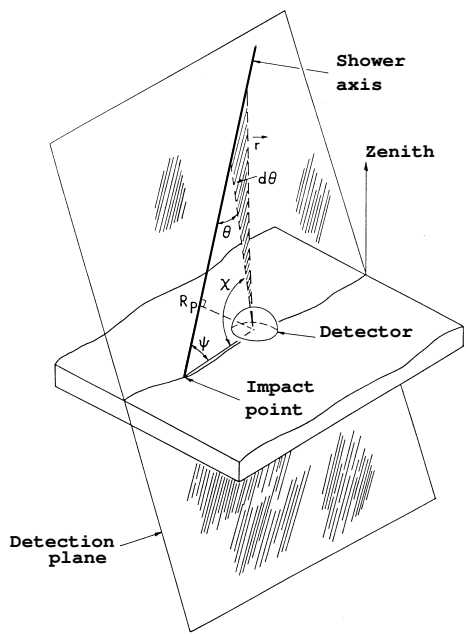} 
\end{center}
\caption{The principle of the detection of an EAS by a fluorescence telescope.\label{fe}} 
\end{figure} 

Figure \ref{fe} shows the geometry of the detection of an air shower by a Fly's
Eye type detectors. The detector sees the shower as a variable light bulb\footnote{A rough estimate of 
the equivalent radiated power would be $3E_{18}$ watts at the shower maximum, where $E_{18}$ is the 
primary energy in EeV.}~ moving at
the speed of light along the shower axis. The detector itself is a set of
phototubes mounted on a ``camera" set at the focal plane of a mirror. Each
phototube sees a small portion of the sky (typically $1^{\circ}$ squared). A fit 
of the hit tubes pattern determines with a 
precision better than one degree the plane containing the detector and the
shower axis. In the \emph{stereo mode} (EAS seen by two telescopes installed a few km apart), 
two planes are thus reconstructed and their intersection gives the incident direction with good
 precision. In the \emph{mono mode} (EAS seen by a single telescope), one relies  on  the 
time of arrival of the photons on the tubes. A good reconstruction of the direction (the $\Psi$ angle)
 then needs a larger number of pixels, enough to measure simultaneously the angular velocity
 and the angular acceleration of the shower development. Finally, in the \emph{hybrid mode}, i.e. 
simultaneous detection of the EAS with a fluorescence telescope \emph{and} a ground array, 
the position of the core given by the array determines the final geometry.
For 100~EeV showers, a precision of $0.2^{\circ}$ can then be reached.

The fluorescence technique is the most appropriate way to measure the
energy of the incident cosmic ray: it is a partial calorimetric measurement with continuous 
longitudinal sampling. The amount of fluorescence light emitted is proportional to the 
number of charged particles in the
shower. The EAS has a longitudinal development usually parameterized by the
analytic Gaisser-Hillas function giving the size $N_{\text{e}}$ of the shower
(actually the number of the ionizing electrons) as a function of atmospheric
depth $x$: 
$$
N_{\text{e}}(x)=N_{\text{max}}\left(\frac{x-x_0}{X_{\text{max}}-x_0}
\right)^{(X_{\text{max}}-x_0)/\lambda}\text{e}^{(X_{\text{max}}-x)/\lambda}
$$
where $\lambda=70\,$g/cm$^2$, $x_0$ is the depth at which the first interaction
occurs, and $X_{\text{max}}$ the position of the shower maximum. The total
energy of the shower is proportional to the integral of this function,
knowing that the average energy loss per particle is 2.2~MeV/g~cm$^{-2}$.

In practice several effects have to be taken into account to properly convert the detected
fluorescence signal into the primary energy. These include the subtraction of the
direct or diffused Cerenkov light, the effects of Rayleigh and Mie
scatterings, the dependence of the attenuation on altitude (and
elevation for a given altitude) and atmospheric conditions, the energy
transported by the neutral particles (neutrinos), the hadrons interacting with 
nuclei (whose energy is not
converted into fluorescence) and penetrating muons whose energy is mostly
dumped into the earth. One also has to take into account that a shower is never seen 
in its totality by a fluorescence telescope: the Gaisser-Hillas function parameters are 
measured by a fit to the visible part of the shower, (there is usually a missing part at the beginning, 
close to the interaction point, and at the end, tail absorbed by the earth). 
All these effects contribute to the systematic errors in
the energy measurement which needs sophisticated monitoring and
calibration techniques. The overall energy resolution one can reach with a
fluorescence telescope is of course dependent on the EAS energy but also on the detection mode (mono, 
stereo or hybrid). The HiRes detector should have a resolution of 25\% or better above 
30~EeV in the \emph{mono} mode.  This improves significantly in the stereo or hybrid modes
 (about 3\% \emph{median} relative error at the same energy in the latter case).

The identification of the primary cosmic ray with a fluorescence telescope is
based on the shower maximum position in the
atmosphere ($X_{\text{max}}$).
Simulations show typical values of 750 and 850
g/cm$^2$ for iron nuclei and protons respectively. 
Unfortunately, the physical fluctuations of the first interaction point and
of the shower development (larger than the measurements precision)
blur this ideal image. 
Therefore, one must look for statistical means of studying the chemical
composition and/or use the hybrid detection method where a multi-variable
analysis becomes possible.

The former method uses the so-called \emph{elongation rate} measured for a
sample of showers within some energy range. The depth of the shower maximum as
a function of the energy for a given composition is given by\cite{Sokolsky}:
$$X_{\text{max}}=D_{\text{el}}\ln\left(\frac{E}{E_0}\right)$$  
where $E_0$ is a parameter depending on the primary nucleus mass. Therefore, 
incident samples of pure composition will be displayed as parallel straight
lines with the same slope $D_{\text{el}}$ (the \emph{elongation rate})
on a semi-logarithmic diagram. 

\subsection{The ground array technique}
\noindent
The surface of the array is a
direct function of the expected incident flux and of the statistics needed to
answer the questions at stake. The 100 km$^2$ AGASA\cite{Yoshoda,Takeda} array is
appropriate to confirm the existence of the UHECR with energies in excess of 100
EeV (which it detects at a rate of about one event per year). To explore the
properties of these cosmic rays and hopefully answer the open question of their
origin, the Auger Observatory with its 6000 km$^2$ surface over two sites will 
be very helpfull.

The array detectors count the number of secondary particles which cross them
as a function of time, sampling the non-absorbed part of the shower which reaches the ground. 
The incident cosmic ray direction and energy are
measured by assuming that the shower has an axial symmetry. This assumption is
valid for not too large zenith angles (usually  $\theta<60^{\circ}$). At larger
angles the low energy  secondaries are deflected by the geomagnetic
fields and the analysis becomes more delicate. 

The direction of the shower axis (hence of the incident primary) is reconstructed by
fitting the ``lateral distribution function" (LDF) to the measured densities. 
The LDF explicit form depends on each
experiment. The Haverah Park experiment\cite{Lawrence} (an array of
water-Cerenkov tanks) used the function:
$$\rho(r,\theta,E)=kr^{-[\eta(\theta,E)+r/4000]}$$
as the LDF for distances less than 1 km from the shower core. Here $r$ is in
meters, and $\eta$ can be expressed as:
$$\eta(\theta,E)=a+b\sec\theta+c\log(E/E_0)$$
with appropriate values for all the parameters taken from shower theory and
Monte Carlo studies in a given energy range. At larger distances (and higher
energies), this function has to be modified to take into account a change in
the rate at which the densities decrease with distance. A much more
complicated form is used by the AGASA group.\cite{Yoshida} However, the
principle remains the same.

Once the zenith angle correction is made for the LDF, an estimator of the
primary energy is extracted from this function. At energies below 10~EeV,
the optimal estimation distance is 600~m from the shower core,
 a value slowly increasing with energy,  reaching 1000m in the UHECR range.
Once this value is determined, the primary energy is related to
it by a quasi-linear relation: 
$$E=k\rho_{optimal}^\alpha$$ 
where $\alpha$ is a parameter close to 1. Of course, to be able to reconstruct
the LDF, many array stations have to be hit at the same time. The
spacing between the stations determines the threshold energy for a vertical
shower: the 500~m spacing of the Haverah Park triggering stations corresponds
to a threshold of a few $10^{16}$~eV, while the 1.5~km separation of the
Auger Observatory stations gives almost  100\% efficiency above 10~EeV.

In a ground array, the primary cosmic ray's identity is reflected in the
proportion of muons among the secondaries at ground level.
Here a proper estimator is therefore the ratio of muons to electrons -~and
eventually photons, if they are detected~-. 
When a ground array has muon detecting capabilities (water
Cerenkov tanks, buried muon detectors), one measures directly the muon to
electron ratio. Otherwise, an indirect method is given by the signal rise time.

\section{Experimental results}
\noindent 
It is outside the scope of this lecture to present the full history of the
cosmic ray detection and studies. This would cover the whole century (1912 is
the year of the first decisive balloon experiments by Victor Hess). As
a starting point for the genesis of the UHECR physics, 
one can use  the first observations of Pierre Auger
 and collaborators\cite{PAuger} done in 1938. They studied the coincidence
rates between counters with increasing separation (up to 150~m in their first
experiments in Paris, more than 300~m when they repeated them at the
Jungfraujoch in Switzerland). They inferred from these very modest measurements
the existence of primary cosmic rays with energies as large as 1~PeV (\E{15}).

\begin{figure}[!htb] 
\begin{center}
\epsfxsize=20pc
\epsfbox{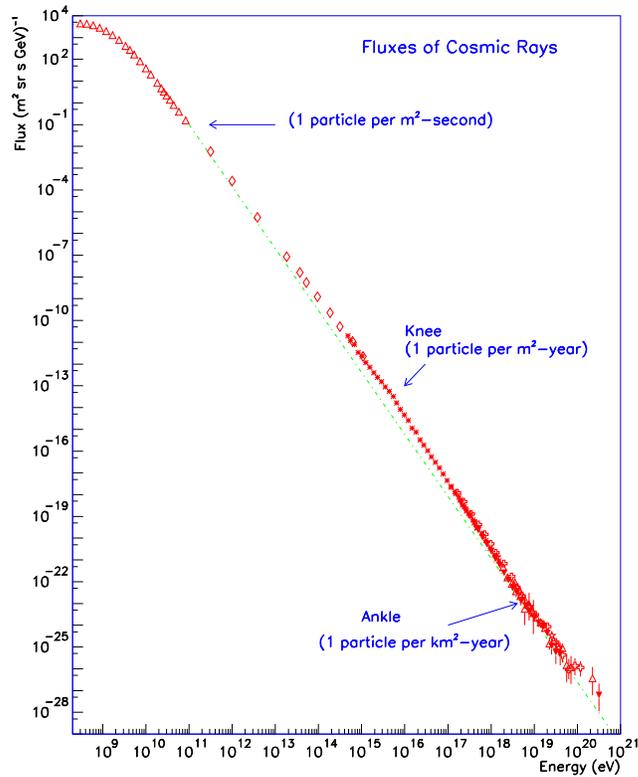}
\end{center}
\caption{The all-particle spectrum of cosmic rays}
\label{spectrum}
\end{figure} 

Figure \ref{spectrum} is a compilation\cite{Swordy} of the differential
spectrum of cosmic ray flux as a function of energy. On this figure, integrated
fluxes above three energy values are also indicated: 1 particle/m$^2$-second
above 1~TeV, 1 particle/m$^2$-year above 10~PeV, 1 particle/km$^2$-year above 10
EeV. Ground detectors are the only alternative for the
highest energy part of the spectrum.

In this section, and unless otherwise specified, we shall pay special attention to the 
events with energies exceeding 100~EeV. This value has no particular
physical meaning except that it is well above the GZK cutoff.

\subsection{The energy spectrum and flux}
\noindent
The energy spectrum (Figure \ref{spectrum}) ranges over 13 orders of magnitude
in energy and 34 orders of magnitude in flux. However, if one discards the
saturation region at the lowest energies, the spectrum is surprisingly regular
in shape. From the GeV energies to the GZK cutoff, it can be represented
simply by three power-law curves 
interrupted by two breaks, the so-called
``knee" and ``ankle". 

\begin{figure}[!htb]
\vspace*{13pt}
\begin{center}  
\mbox{ \epsfig{file=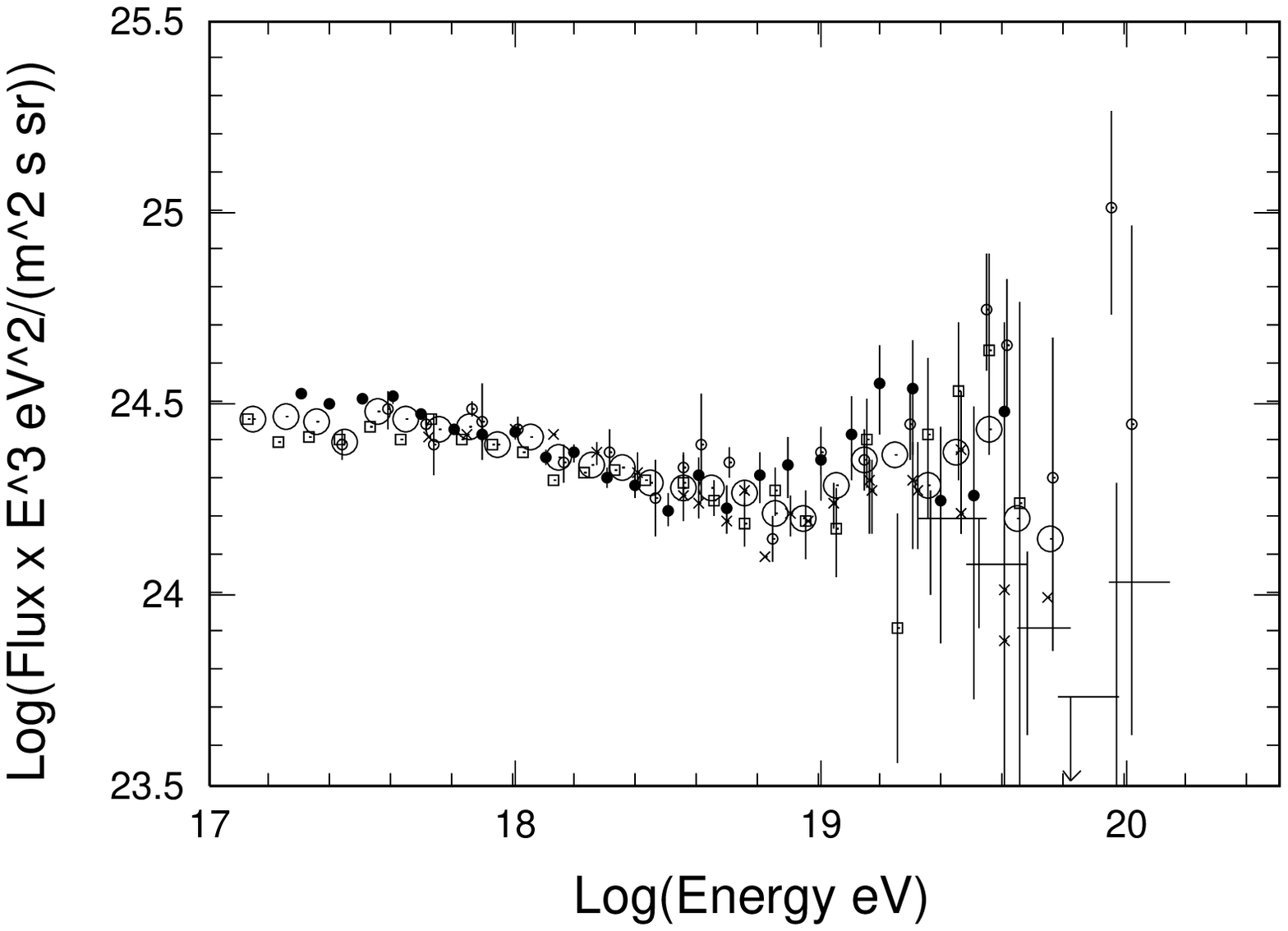,width=270pt}}
\end{center}  
\vspace*{13pt}
\caption{Energy spectrum above 100~PeV. The compilation is from
Ref. updated by M.Nagano (private communication).\label{agasa}} 
\end{figure}

\begin{figure}[!htb]
\vspace*{13pt}
\begin{center}  
\epsfig{file=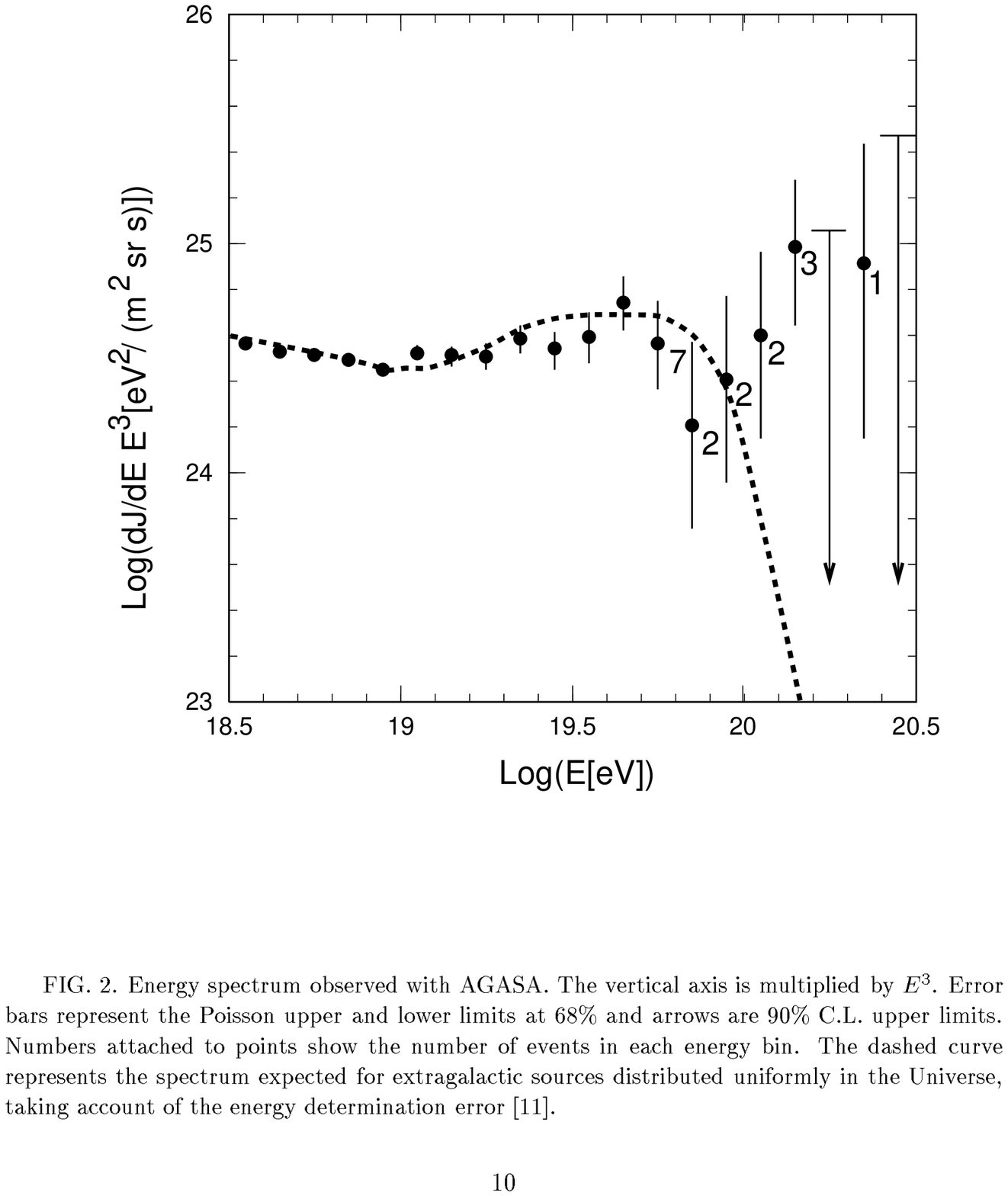,bbllx=117pt,bblly=228pt,bburx=512pt,%
bbury=602pt,width=8cm,clip=}
\end{center}  
\vspace*{13pt}
\caption{Highest energy region of the cosmic ray spectrum as observed by the
AGASA detector The figures near the data points indicate the
number of events in the corresponding energy bin. The arrows show 90\%
confidence level upper limits. The dashed line is the expected spectrum if the
sources were cosmologically distributed.\label{takeda}} 
\end{figure}

The flux of supra-GZK events is extremely low. Figure \ref{agasa} is a zoom on the
highest energy part of the total spectrum. On this figure, 
the energy spectrum is multiplied by $E^3$ so
that the part below the EeV energies becomes flat. One can see the `ankle'
structure in its complexity: a steepening around the EeV and then a confused
region where the GZK cutoff is expected. The ultimate data points come from
very few events hence their large error bars. Due to normalization problems it is
difficult to compare different experiments. On Figure \ref{takeda} where the
AGASA data alone are displayed,\cite{Takeda} one has a clearer view of what can
be expected from a cosmological distribution of conventional 
sources and what is observed. 

The GZK cutoff is clearly visible on the dashed line while the data suggests 
a change of slope as if a new phenomenon was rising above a steeply falling spectrum.
The cutoff, that would be expected if the sources were cosmologically
distributed and if the observed cosmic rays had no exotic propagation or interaction 
properties, \emph{is} not present in the observed data.

The flux of the highest energy cosmic rays cannot be deduced from the
data above the cutoff energies: no reliable fit to the spectrum shape is
feasible in this region. A reasonable estimate can be ontained taking the ratio
of the total experimental exposure to the number of events observed. 
The exposure to events above the GZK cutoff for AGASA, Fly's Eye and Haverah
Park detectors together, is of the order of 2000
km$^2$~sr~yr. The number of events observed in this energy range
 yields an integrated flux which can be parameterized by\label{flux}:

\begin{equation}
I(E>E_0)\approx\left(\frac{E_0}{10\,\text{EeV}}\right)^{-2}\,
\text{km}^{-2}\text{sr}^{-1}\text{year}^{-1}\label{eq:flux}
\end{equation} 
 With $E_0=100$ EeV. Therefore the expected flux above 100 EeV is 
(only) 1 particle per km$^2$ per century.

\subsection{The chemical composition}
\label{Compo}
\noindent
The UHECR chemical composition is very likely to be
unveiled only on a statistical basis. What we know at present 
is weak and controversial due to the limited number of events observed. 
The most
recent information comes from the Fly's Eye and AGASA experiments. The  Fly's
Eye studies\cite{Bird} (between 0.1~EeV and 10~EeV) are based on $X_{\text{max}}$ 
behavior as a function of the logarithm of the primary energy.  With this
method, their data show evidence of a shift from a dominantly heavy composition
(compatible with iron nuclei) to a light composition (protons). 
In this framework UHECR are mainly protons.

The AGASA group based their primary identification on the
muon content of the EAS at ground level,\cite{Hayashida}.
Initially the conclusion of the AGASA experiment was quite opposite to the Fly's Eye: no
change in chemical composition. However a recent critical review of both
methods\cite{Dawson} showed that the inconsistencies were mainly due to the
scaling assumption of the interaction model used by the AGASA group. 
The authors concluded that if a model with a higher (compared to the one given by scaling) 
rate of energy dissipation at high energy is assumed, as indicated by the direct $X_{\text{max}}$ 
measurements of the Fly's Eye, both data sets demonstrate a change of composition, a shift 
from heavy (iron) at 0.1~EeV to light (proton) at 10~EeV. Different interaction models as long
as they go beyond scaling in their energy dissipation, would lead to the same qualitative 
result but possibly with a different rate of change.

Gamma rays have also high cross sections with air and are still another possible candidate 
for  UHECR but no evidence were found up to now for a gamma signature among the Big Events. 
The most energetic Fly's Eye event was studied in detail\cite{Halzen2} and found incompatible with
an electromagnetic shower. Both interpretations of the AGASA and the Fly's Eye data 
favor a hadronic origin.

\subsection{Distribution of the sources} 
\noindent 
A necessary ingredient in the search for the origin of the UHECR is to locate
their sources. This is done by reconstructing the incident cosmic ray's
direction and checking if the data show images of point sources or correlations
with distributions of astrophysical objects in our vicinity. In the following we will
consider the effects of the galactic and extragalactic magnetic
fields on protons and we will show that for supra-GZK energies, proton astronomy is
possible to some extent. We will then give a review of what we can extract from
the present data.

\subsubsection{Magnetic fields}\label{MagField} 
\noindent 
There are a limited number of methods to study the magnetic fields on 
galactic or extragalactic scales.\cite{Kronberg} One is the measure of the
Zeeman splitting of radio or maser lines in the interstellar gas. This method
informs us mainly on the galactic magnetic fields, as extragalactic signals suffer Doppler smearing
while the  field values are at least three orders of
magnitude below the galactic ones. The 
magnetic field structure of the galactic disc is therefore thought to be rather well
understood. One of the parameterizations currently used is that of Vall\'ee\cite{Vallee}: concentric
field lines with a few $\mu$G strength and a field reversal at about one half of
the disk radius. Outside the disk and in the halo, the field model is based on
theoretical prejudice and represented by rapidly decreasing functions (e.g. gaussian
tails).

The study of extragalactic fields is mainly based on the Faraday rotation
measure (FRM) of the linearly polarized radio sources. The rotation angle
is a measurement of the integral of $n_eB_{\parallel}$ along the line of sight, 
where $n_e$ is the electron/positron density and $B_{\parallel}$ the
longitudinal field component. Therefore, the FRM
needs to be complemented by the measurement of $n_e$. 
This is done
by observing the relative time delay versus frequency of waves emitted by a
pulsar. Since the group velocity of the signal depends simultaneously on its
frequency but also on the plasma frequency of the propagation medium,
measurement of the dispersion of the observed signals gives an upper limit on
the average density of electrons in the line of sight. 
Here again, because of the faintness of extragalactic signals, our knowledge of
the strength and coherence distances of large scale extragalactic fields is
quite weak and only upper limits over large distances can be extracted. An
educated guess gives an upper limit of 1 nG for the field
strength and coherence lengths of the order of 1 Mpc.\cite{Kronberg} 

\begin{figure}[htbp]
\vspace*{13pt}
\begin{center}  
\epsfig{file=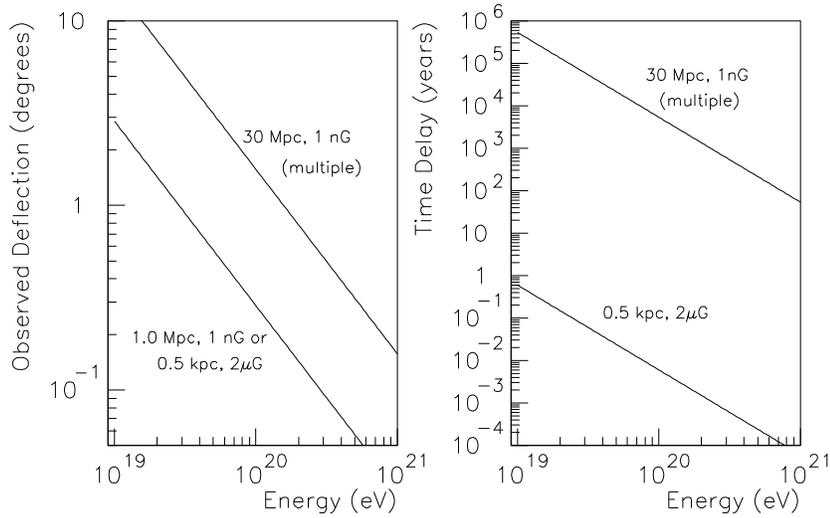,height=214pt,width=322pt}
\end{center}  
\vspace*{13pt}
\caption{Effect of magnetic fields on the propagation of a proton as a
function of energy: angular deviation (left) and time delay (right), with
respect to a straight line trajectory, in the framework of three realistic
scenarios (see text).\label{angular}} 
\end{figure}
A few other more or less indirect methods exist for the study of large scale
magnetic fields. \emph{If} the UHECR are protons and \emph{if} they come from
point-like sources, the shape of the source image as a function of the cosmic
ray energy will certainly be one of the most powerful of them. The Larmor radius
$R$ of a charged particle of charge $Ze$ in kiloparsecs is given by:
$$R_{\text{kpc}}\approx\frac{1}{Z}\left(\frac{E}{1\,\text{EeV}}\right)
\left(\frac{B}{1\,\mu\text{G}}\right)^{-1}$$
The Larmor radius of a charged particle at 320~EeV is larger than the size of
the galaxy if its charge is less than 8. If we take the currently accepted
upper limit ($10^{-9}$~G) for the extragalactic magnetic fields, a proton of
the same energy should have a Larmor radius of 300~Mpc or more.

In Figure \ref{angular}, three different situations are envisaged to evaluate
the effects of magnetic fields on a high energy cosmic proton. The situations
correspond to what is expected a-) for a trajectory through our galactic disk
(0.5~kpc distance inside a 2~$\mu$G field) or b-) over a short distance (1~Mpc)
through the extragalactic (1~nG) field (same curve), and finally c-) a 30~Mpc
trajectory through extragalactic fields with a 1~Mpc coherence length (multiple
scattering effect). One can see that at 100~EeV, the deviation in the third
case would be about $2^{\circ}$. This gives an idea of the image size if the
source is situated inside our local cluster or super-cluster of galaxies.
Since the angular resolution of the (present and future) cosmic ray detectors
can be comparable to or much better than this value, we expect to be able to
locate point-like sources or establish correlations with
large-scale structures. 

However, let us remember that this working hypothesis of very weak extragalactic
magnetic fields is not universally accepted. Several authors recently advocated
our bad knowledge of those fields arguing for  stronger magnetic fields (typically at the
$\mu$G level) either locally\cite{Lemoine} or distributed over larger,
cosmological, scales.\cite{Farrar2}\label{Xfield}

\subsubsection{Anisotropies}
\noindent
\begin{figure}[!htb]
\vspace*{13pt}
\begin{center}  
\epsfig{file=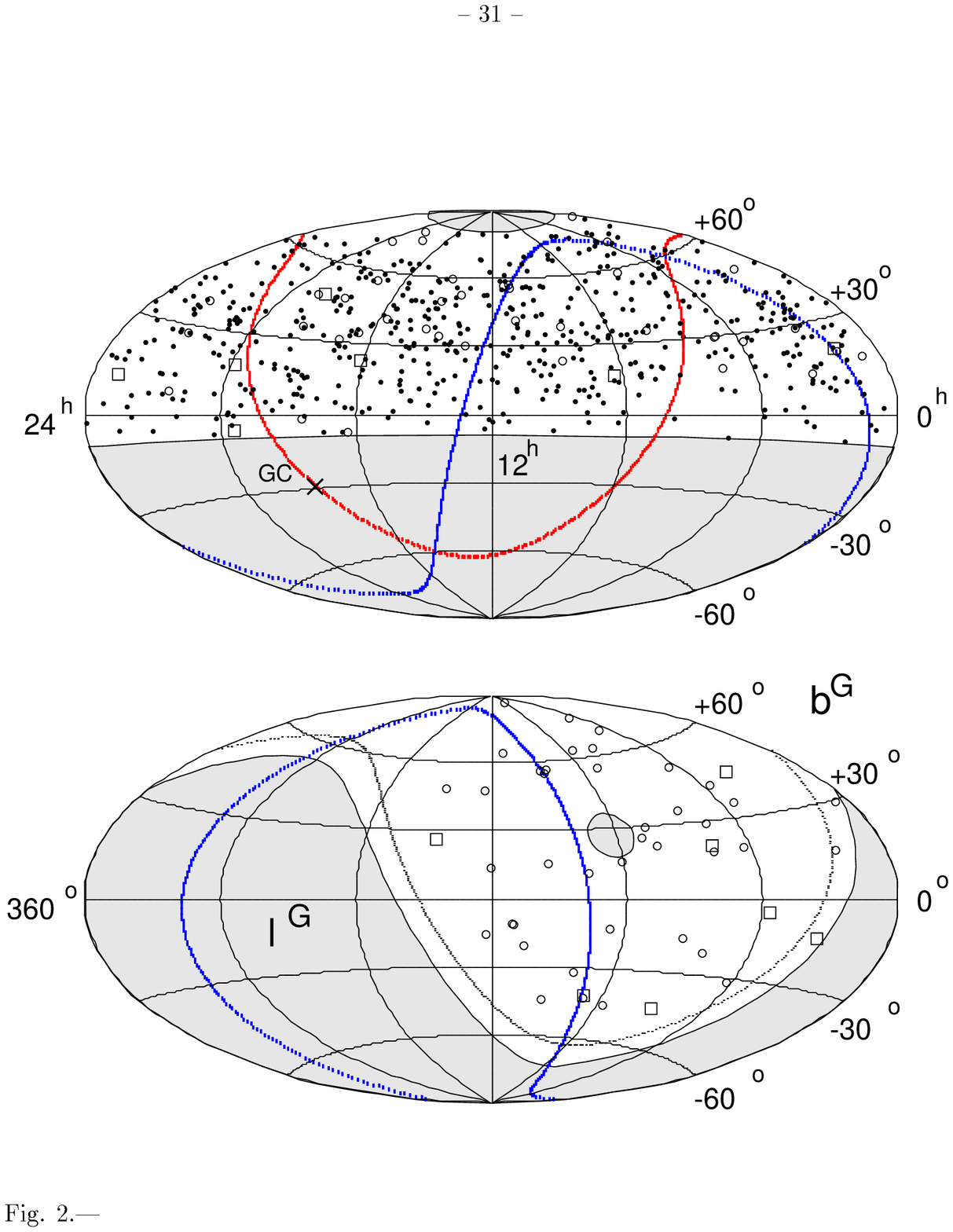,bbllx=79pt,bblly=417pt,bburx=528pt,%
bbury=627pt,width=12cm,clip=}
\end{center}  
\vspace*{13pt}
\caption{Arrival directions of cosmic rays with energies above 10~EeV
(equatorial coordinates), as measured by the AGASA experiment.
The thick dotted lines show the galactic and supergalactic planes (GC
indicating the galactic center). The shaded regions are those invisible to the
AGASA detector. See text.\label{bigevents1}}
\end{figure}
In the search for potential sources, the propagation arguments incite us to
look for correlations with the distribution of astrophysical matter within a
few tens of Mpc. In our neighborhood, there are two structures
showing an accumulation of objects, both only partially visible from any
hemisphere: the galactic disk on a small scale and the supergalactic plane on a large scale, a
structure roughly normal to the galactic plane, extending to distances up to
$z\approx 0.02$ (about 100 Mpc).

In equatorial coordinates isotropically distributed sources, 
give a uniform right ascension distribution of events
and a declination distribution which can be parameterized with the
known zenith angle dependence of the detector aperture. 

The most recent analysis on the correlation between arrival directions and
possible source locations was done by the AGASA experiment for the highest 
energy range.\cite{Takeda2} The
analysis is based on 581 events above 10~EeV, a subset of 47 events above 40
EeV and 7 above 100~EeV. Figure \ref{bigevents1} is a compilation of the total sample in equatorial
coordinates. The dots, circles and squares are respectively events with
energies above 10, 40 and 100~EeV. The data show no
deviation from the expected uniform right ascension distribution. An excess of
2.5~$\sigma$ is found at a declination of $35^{\circ}$ and can be interpreted as
a result of observed clusters of events (see below). No convincing deviation
from isotropy is found when the analysis is performed in galactic coordinates.

The same collaboration\cite{Hayashida2} made also a similar analysis for the lower energy region (events
down to 1~EeV and detected with zenith angles up to $60^{\circ}$). In this
article, a slight effect of excess events in the direction of the galactic
center was announced. A similar study\cite{Bird2} with the Fly's Eye data, concludes on a small
correlation with the galactic plane for events with energies lower than 3.2
EeV and isotropy for higher energies.

In summary, both the AGASA and Fly's Eye experiments seem to converge on some anisotropy in the EeV range (correlation with the galactic plane and center) and isotropy above a few tens of EeV. This result may seem surprising - one
naively expects the correlations to be stronger when the cosmic rays have large
magnetic rigidity. It is actually explained by the fact that the low energy
component may be dominantly galactic heavy nuclei (see the section on chemical
composition), hence a (weak) correlation with the galactic plane, whereas the
higher energy cosmic rays would be dominated by extragalactic protons.

\subsection{Point sources?} 
\noindent If the sources of UHECR are nearby astrophysical objects and if, as
expected, they are in small numbers,
a selection of the events with the largest magnetic rigidity would combine
into multiplets or clusters which would indicate the
direction to look for an optical or radio counterpart. Such
an analysis was done systematically by the AGASA group.\cite{Takeda2}  
\begin{figure}[!htb] 
\vspace*{13pt}
\begin{center}  
\epsfig{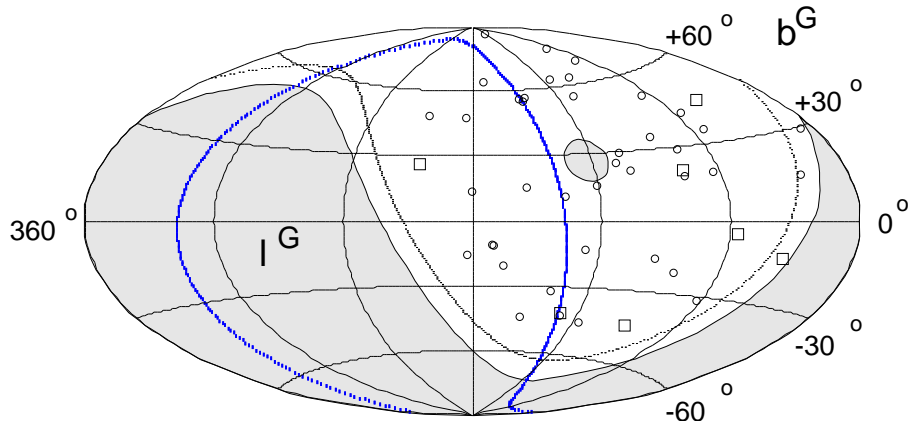} 
\end{center}   
\vspace*{13pt} 
\caption{Arrival directions of cosmic rays with energies above 40~EeV
(galactic coordinates), as measured by the AGASA experiment.See
text.\label{bigevents2}} 
\end{figure} 
Figure \ref{bigevents2} shows the subsample of events in the AGASA catalog with
energies in excess of 100~EeV (squares) and in the range 40-100~EeV (circles).
A multiplet is defined as a group of events whose error boxes ($2.5^{\circ}$
circles) overlap. One can see that there are three doublets and one triplet. If
one adds the Haverah Park events, the most southern doublet also becomes a
triplet. The chance probability of having as many multiplets as observed with
a uniform distribution are estimated by the authors to less than 1\%.\footnote{The chance probability is very difficult to evaluate in an \emph{a posteriori} analysis and depends 
strongly on the assumed experimental error box size.}

A search for nearby astrophysical objects within an angle of $4^{\circ}$ from any event
in a multiplet was also done, and produced a few objects. One of the most
interesting candidates is Mrk 40, a galaxy collision, since the shock
waves generated in such phenomena are considered by some authors\cite{Cesarsky}
as being valid accelerating sites.

Another way of using the observed multiplets, \emph{assuming} that they come
from an extragalactic point source, is to consider the galactic disk as a 
magnetic spectrometer which can give information on the charge of the incident
cosmic rays. Cronin\cite{Cronin} made such an analysis on the doublet where
the energy difference between the two events is the largest (a factor of four).
He uses the magnetic field model of Vall\'ee\cite{Vallee} to trace back the
detected couple of events outside of the galaxy assuming various charges. It
is shown that the maximum charge compatible with a separation less than the
detector's angular resolution is 2 for the members of the doublet with
conservative integrated values for the magnetic field, a result 
compatible with UHECR being mostly protons.

\section{Conclusions}
\noindent
The UHECR were a puzzle when they were first observed, more than 30 years ago. 
They still are. Among all the tentative explanations given to their existence
none fully explains the whole set of observation.

The past experiments which explored this field could do hardly better than 
convince us of the existence of the UHECR above the GZK cutoff. Statistics 
which should make us able to locate the sources, reconstruct the shape of the
cosmic ray spectrum above the cutoff and study the  chemical composition will  soon
be provided by the ongoing (HiRes, AGASA) and oncoming (Auger,
Telescope-Array,  OWL/Airwatch, EUSO) experiments.
\vspace*{1.2cm}
\par
\noindent
The references given as astro-ph/xxxxxxx or hep-ph/xxxxxxx are articles
available from the Web electronic preprint archive at the URL {\tt http://xxx.lanl.gov/}


\begin{thebibliography}{000}
\bibitem{Hemas} Web site: {\tt osf1.lngs.infn.it/\lower
3pt\hbox{\string~}carboni/Hemas/doc/}
\end{thebibliography}
\end{document}